\documentclass[12pt,usenames,dvipsnames,a4paper]{article}

\usepackage{changepage}
\usepackage{amsmath,amsfonts,amssymb}
\usepackage{epsf,amsmath,bbold,amsfonts,stmaryrd}
\usepackage{mathrsfs}
\usepackage{appendix}
\usepackage{caption}
\usepackage{cite}
\usepackage{color}
\usepackage{datetime}
\usepackage{float}
\usepackage{graphicx}
\usepackage[colorlinks]{hyperref}
\hypersetup{pageanchor=false,citecolor=red,urlcolor=red}
\usepackage{indentfirst}
\usepackage[numbers,square,comma,sort&compress,merge]{natbib} 
\usepackage{subfig}
\numberwithin{equation}{section}
\usepackage{mathtools}
\usepackage[colorinlistoftodos]{todonotes}
\usepackage{ytableau}
\usepackage{tabu}

\allowdisplaybreaks

\def\a{\alpha}

\def\b{\beta}

\def\const{\mathrm{const}}

\def\d{\delta}
\def\D{\Delta}

\def\eps{\varepsilon}
\def\f{\frac}

\def\l{\left}

\def\la{\langle}
\def\ra{\rangle}

\def\mc{\mathcal}

\def\m{\mu}
\def\n{\nu}

\def\nn{\nonumber}

\def\p{\partial}
\def\vp{\varphi}

\def\r{\right}
\def\s{\sigma}
\def\t{\theta}

\def\vp{\varphi}

\def\x{\xi}

\def\be{\begin{equation}}
\def\ee{\end{equation}}

\def\bea{\begin{eqnarray}}
\def\eea{\end{eqnarray}}

\def\ba{\begin{array}}
\def\ea{\end{array}}

\def\bc{\begin{center}}
\def\ec{\end{center}}

\def\bl{\begin{flushleft}}
\def\el{\end{flushleft}}

\def\br{\begin{flushright}}
\def\er{\end{flushright}}

\def\bi{\begin{itemize}}
\def\ei{\end{itemize}}

\def\bt{\begin{tabular}}
\def\et{\end{tabular}}

\newcommand{\REF}[1]{(\ref{#1})}

\makeatletter

\newsavebox\myboxA
\newsavebox\myboxB
\newlength\mylenA

\newcommand*\xoverline[2][0.75]{%
    \sbox{\myboxA}{$\m@th#2$}%
    \setbox\myboxB\null
    \ht\myboxB=\ht\myboxA%
    \dp\myboxB=\dp\myboxA%
    \wd\myboxB=#1\wd\myboxA
    \sbox\myboxB{$\m@th\overline{\copy\myboxB}$}
    \setlength\mylenA{\the\wd\myboxA}
    \addtolength\mylenA{-\the\wd\myboxB}%
    \ifdim\wd\myboxB<\wd\myboxA%
       \rlap{\hskip 0.5\mylenA\usebox\myboxB}{\usebox\myboxA}%
    \else
        \hskip -0.5\mylenA\rlap{\usebox\myboxA}{\hskip 0.5\mylenA\usebox\myboxB}%
    \fi}
\makeatother

\begin{document}
\allowdisplaybreaks

\vspace*{-2.5cm}
\begin{adjustwidth}{}{-.45cm}
\br
{
\begin{tabular}{@{}l@{}}
\small LMU--ASC~39/21
 \end{tabular}
 }
\er
\end{adjustwidth}

\vspace*{1.5cm}
\begin{adjustwidth}{-1.3cm}{-.7cm}

\begin{center}
    \bf \Large{More on the operator-state map in non-relativistic CFTs}
\end{center}
\end{adjustwidth}

\begin{center}
\textsc{Georgios K. Karananas,$^\star$~Alexander Monin\,$^{\dagger,\ddagger}$}
\end{center}

\begin{center}
\it {$^\star$Arnold Sommerfeld Center\\
Ludwig-Maximilians-Universit\"at M\"unchen\\
Theresienstra{\ss}e 37, 80333 M\"unchen, Germany\\
\vspace{.4cm}
$^\dagger$Institute of Physics \\
Theoretical Particle Physics Laboratory (LPTP) \\ 
\'Ecole Polytechnique F\'ed\'erale de Lausanne (EPFL) \\ 
CH-1015 Lausanne, Switzerland\\
\vspace{.4cm}
$^\ddagger$Department of Physics and Astronomy\\ University of South Carolina\\
Columbia SC 29208, USA
}
\end{center}

\begin{center}
\small
\texttt{\small georgios.karananas@physik.uni-muenchen.de}  \\
\texttt{\small alexander.monin@unige.ch} 
\end{center}

\begin{abstract}

We propose an algebraic construction of the operator-state correspondence in non-relativistic conformal field theories by explicitly constructing an automorphism of the Schr\"odinger algebra relating generators in different frames. It is shown that the construction follows closely that of relativistic conformal field theories.

\end{abstract}

\section{Introduction}

The non-relativistic conformal group is the symmetry group of the free Schr\"odinger equation.\footnote{One may argue that it is only natural to take as the non-relativistic analogue of the conformal group the transformations commuting with the free Schr\"odinger equation, in view of the fact that the conformal group corresponds to the symmetries of the free massless Klein-Gordon operator.} On top of the Galilei subgroup, it includes non-relativistic dilatations and one special conformal transformation. Theories invariant under the (centrally extended) Schr\"odinger group are called non-relativistic conformal field theories (NRCFTs). Examples of those are non-relativistic particles with an $r^{-2}$ potential interaction and fermions at unitarity~\cite{Son:2005rv,Nishida:2007pj,Son:2008ye}.

The name NRCFT can be somewhat misleading, for it evokes conformal field theory (CFT) suggesting that the former is a special case of the latter. However, NRCFT is only (if at all) a distant cousin of CFT. The Schr\"odinger group cannot be obtained (at least in the same number of dimensions~\cite{Son:2008ye}) from the conformal group by considering the non-relativistic limit. In other words, the relation between the two groups is not the same as between the Poincar\'e and Galilei groups, where the latter is the In\"on\"u-Wigner contraction of the former.\footnote{Actually, by performing the contraction of the conformal algebra one ends up with yet another type of nonrelativistic conformal algebra. This has the same number of generators as its parent. For more details see~\cite{Bagchi:2009my}.}
 
Even putting aside the central charge $Q$ corresponding to the
particle number in non-relativistic theory, the number of generators
for the two groups is different: CFT has as many special conformal
generators as the number of space-time dimensions, while in NRCFT
there is only one analogue of the special conformal transformation,
irrespectively of the spacetime dimensionality. Dilatations are also
different in the two theories, since in CFT these do not distinguish
between space and time, while in NRCFT time and spatial coordinates
scale differently; this in turn allows to have dimensionful parameters, such as mass, clearly forbidden in CFT.
 
From an effective field theory perspective, the symmetry group of NRCFT is an accident of the non-relativistic limit. Considering higher order (in inverse powers of the speed of light) operators would bring in symmetry-breaking terms revealing that NRCFT originates from a Poincar\'e invariant theory rather than from a theory with an enhanced symmetry like a CFT.
 
Despite all the differences there are common features of conformal field theories and their non-relativistic counterparts, allowing to draw general conclusions about the two types of theories. For instance, the operators in both theories are organized into primaries and descendants and the operator product expansions (OPE) are determined by the corresponding contributions from primary operators. It can be shown~\cite{Golkar:2014mwa,Goldberger:2014hca} that in both theories OPE has a finite radius of convergence and, which is intimately related to this fact, that both types of theories posses what is called an operator-state correspondence. The latter establishes a one-to-one map between states in the Hilbert space and the operators of a theory. In particular, the scaling dimensions of the operators are given by the energies of the corresponding states. One of the consequences of the operator-state correspondence is the presence of unitarity bounds on the scaling dimensions.

The fact that OPE converges also implies that higher order correlators can be expressed in terms of two point functions by applying the OPE  repeatedly, in the same manner as in CFTs. However, developing a bootstrap program for NRCFT is complicated by the fact that three point functions are not fixed completely by kinematics only, as opposed to CFT (see~\cite{Goldberger:2014hca} for more details). 

Our main objective in this paper is to put forward an algebraic construction of the operator-state correspondence for NRCFTs, which parallels the procedure used when studying CFTs. Specifically, we discuss how the Hilbert space structure is introduced on the space of Euclidean fields. This is achieved by finding an automorphism relating the Minkowski and Euclidean generators of the conformal group. This way, the operator-state map is a natural aftermath of the proposed construction. 
 
This work is organized as follows. In Sec.~\ref{sec:o-s_CFT} we set
the stage by rephrasing some well known results of CFTs in a language
which can be used almost verbatim in the NRCFTs. We first give a brief
overview of some basics about the conformal algebra and its unitary
representations. Then, we turn to the operator-state correspondence
and how this emerges as a consequence of the mapping between the
Minkowski and Euclidean generators of the conformal algebra and the
operator algebra. Sec.~\ref{sec:NRCFT} is devoted to NRCFTs. Namely, we introduce the
Schr\"odinger group and discuss the algebra's representations and
action on operators. Then, we construct the appropriate map between
the generators of the algebra by finding the corresponding coordinate
transformation. Finally, we explicitly demonstrate that in the
non-relativistic considerations, confining the theory in a harmonic
trap is completely analogous to putting a CFT on the cylinder. We
conclude in Sec.~\ref{sec:conclusions}. Various
technical details can be found in the Appendices. 

\section{CFTs}
\label{sec:o-s_CFT}

\subsection{The conformal algebra}

We start our discussion by considering the conformal algebra in a  $d$-dimensional Minkowski spacetime. The commutation relations between the generators of translations $P_\m$, Lorentz transformations $J_{\m \n}$, dilatations $D$, and special conformal transformations $K_\m$, read
\be
\begin{aligned}
\l [ D, P _ \m \r ] & =  - i P _ \m \ ,  \\
\l [ J _ {\m\n}, P _ \rho \r ] & =  i \l ( \eta _ {\n\rho} P _ \m -
\eta _ {\m\rho} P _ \n \r ) \ ,  \\
\l [ K _ \m, P _ \n \r ] & =  - 2 i \l ( \eta _ {\m\n} D + J _ {\m\n}
\r ) \ ,  \\ 
\l [ D, K _ \m \r ] & =  i K _ \m\ , \\
\l [ J _ {\m\n}, J _ {\rho\s} \r ] & =  i \l ( J _ {\m\s} \eta _
{\n\rho} + J _ {\n\rho} \eta _ {\m\s} -
J _ {\n\s} \eta _ {\m\rho} - J _ {\m\rho} \eta _ {\n\s} \r ) \ ,  \\
\l [ J _ {\m\n}, K _ \rho \r ] & =  i \l ( \eta _ {\n\rho} K _ \m -
\eta _ {\m\rho} K _ \n \r ) \ ,
\end{aligned}
\label{eq:AlgebraM}
\ee
where 
\be
\eta_{\m\n} = \mathrm{diag}(+,-,\dots,-,-) \ ,~~~\m,\n=0,\ldots,d-1 \  , 
\ee
is the (mostly minus) $d$--dimensional Minkowski metric.  

It is well known that the conformal algebra is equivalent to the algebra of $SO(2,d)$ acting in a $d+2$ dimensional space endowed with metric
\be
\eta _ {AB} = \mathrm{diag}(+,-,\dots,-,+) \ ,~~~A,B=0,\ldots,d+1 \ . 
\ee
To see this explicitly, it is convenient to introduce the following linear combinations of generators\,\footnote{By
construction $M _ {AB}= - M _ {BA}$.}
\be
\label{eq:gener_matrix}
M_{AB} =
\begin{pmatrix}
  J_{\m\n} & -  \f {1} {2} \l(P _ \m -  K _ \m \r )  & -
  \f {1} {2}\l(P _ \m + K _ \m \r )\\
  \f {1} {2} \l(P _ \m -  K _ \m \r )& 0 & -D \\
  \f {1} {2}\l( P _ \m + K _ \m \r )&  D& 0 
\end{pmatrix} \ ,
\ee
or in other words
\be
M_{\m \n} = J _ {\m \n} \ ,~~~M _ {d, \m} = \f {1} {2} \l(P _ \m -  K
_ \m \r ) \ ,~~~M _ {d+1,\m} =
\f {1} {2}\l( P _ \m + K _ \m \r ) \ ,~~~M _ {d+1,d} = D \ .
\label{eq:MvsPKDJ}
\ee 
Using the commutation relations~\REF{eq:AlgebraM} it is straightforward to show that the $M_{AB}$'s indeed satisfy a Lorentz algebra
\be
\l [ M _ {AB}, M _ {CD} \r ] = i \l ( M _ {AD} \eta _ {BC} + M _ {BC}
\eta _ {AD} - M _ {BD} \eta _ {AC} - M _ {AC} \eta _ {BD} \r ) \ .
\label{mink_conf_cr}
\ee

\subsection{Unitary representations of the conformal algebra}
\label{sec:unit_reps_conf}

The unitary representations of $SO(2,d)$ (for which $M_{AB}^\dagger=M_{AB}$) are built by considering its largest compact subgroup which is $SO(2)\times SO(d)$. These correspond to rotations in the $(0,d+1)$-- and $(a,b)$-- planes, respectively; here, $a,b=1,\dots,d$.  The Cartan generators of $SO(d)$ and $M_{d+1,0}$ can be diagonalized simultaneously, therefore, any state in the Hilbert space can be labelled by their eigenvalues. For instance in $d=3$, every vector $| h, l, m \ra$ has the following properties~\cite{Nicolai:1984hb}\,\footnote{For different dimensions see e.g.~\cite{Vichi:2011zza}.} 
\bea
\label{eq:CFTRepresentationCartan}
M_{d+1,0} | h, l, m \ra & = & h | h, l, m \ra \ ,  \nn \\
M_{12} | h, l, m \ra & = & m | h, l, m \ra \ , \\
M_{ab} M_{ab} | h, l, m \ra & = & l(l+1) | h, l, m \ra \ .\nn
\eea
Let us introduce 
\be
M^{\pm} _a=M_{d+1,a} \pm i M_{a,0} \ ,~~~\l ( M^\pm_a \r )^\dagger = M^\mp_a \ . 
\label{eq:RaisingLoweringGenerators}
\ee
It is straightforward to show that 
\be
[M_{d+1,0}, M_a^\pm] = \pm M_a^{\pm} \ ,
\ee
meaning that the generators $M_a^{\pm}$ act as raising and lowering operators for $M_{d+1,0}$
\be
M_{d+1,0} M_a^{\pm} | h, l, m \ra = \l ( h \pm 1 \r ) M_a^{\pm} | h, l, m \ra \ .
\ee
Introducing the lowest weight vector $| h_0 , l_0, m \ra$, for which
\be
\label{eq:CFTRepresentationLowest}
M^-_{a} | h_0, l_0, m \ra = 0 \ , 
\ee
allows to define a representation generated by the raising operators $M_a^+$. 

\subsection{Operator-state correspondence and OPE}
\label{sec:oper_stat_cor}

States in the so-constructed Hilbert space are in one-to-one correspondence with fields in the theory. To make this point clear, let us consider a field, say $\phi(x)$, that is inert under special conformal transformations at the origin $x=0$, i.e. a~\emph{primary} field. We also take it to belong to an irreducible representation of the Lorentz group.  Then, the action of the conformal algebra on $\phi(x)$ can be constructed as a representation induced from that of the stability subalgebra generated by $J_{\m \n}$,  $D$ and $K_\m$. It is straightforward to show that~\cite{Mack:1969rr,Ferrara:1973yt} (see also~\cite{DiFrancesco:1997nk,Rychkov:2016iqz})
\bea
\l [ P _ \m, \phi (x) \r ] & = & - i \p _ \m \phi (x) \ , \nn \\
\l [ M _ {\m \n}, \phi (x) \r ] & = &  i \l ( \Sigma _ {\m \n} - x _
\m \p _ \n + x _ \n \p _ \m \r ) \phi (x) \ ,  \\
\l [ D, \phi (x) \r ] & = & - i \l ( \Delta + x ^ \m \p _ \m \r ) \phi 
(x) \ , \nn \\ 
\l [ K _ \m, \phi (x) \r ] & = & -i \l ( 2 x _ \m x ^ \n - \d ^ \n _
\m x ^ 2 \r ) \p _ \n \phi (x) - 2 i \l ( x _ \m \Delta - x ^ \n
\Sigma _ {\m \n} \r ) \phi (x) \ , \nn
\label{conf_group_transf}
\eea
where $\Sigma_{\m \n}$ corresponds to a finite dimensional (therefore, non-unitary) representation of the Lorentz $SO(d-1,1)$ group and $\D$ is the scaling dimension. In the above, summation over repeated indices is assumed. It follows that at $x=0$ the conformal algebra acts on primary fields as 
$$
\l [ P _ \m, \phi (0) \r ] = - i \p _ \m \phi (0) \ ,~~~\l [ M _ {\m
  \n}, \phi (0) \r ] =  i \Sigma _ {\m \n} \phi (0) \ ,~~~\l [ D, \phi
(0) \r ] = - i \Delta \phi (0) \ ,~~~\l [ K _ \m, \phi (0) \r ] = 0 \
, 
$$
implying an analogy between the sets of generators $\l\{M_a^+\r.$,
$M_{\m \n}$, $M_{d+1,0}$, $\l.M_a^-\r\}$ acting on the Hilbert space
(see Sec.~\ref{sec:unit_reps_conf}) and $\l\{P_\m\r.$, $J_{\m \n}$, $D$, $\l.K_\m\r\}$ acting on fields. Therefore, given an automorphism (modulo an analytic continuation) mapping one set of generators onto the other, the unitary representation discussed above can be viewed as the action of the conformal algebra on fields, where the lowest weight vectors are nothing else but primary operators.

Since we are interested in finite-dimensional representations of $SO(d)$ generated by $M_{ab}$, the action on fields will be consistent with unitarity only if we consider the Euclidean conformal algebra. Indeed, for the Euclidean version,  the corresponding matrix $\Sigma$ need not be infinite-dimensional without contradicting unitarity. Similarly, it is clear that the new generator of dilatations should be identified with $-i M_{d+1,0}$, so it is an anti-Hermitian operator. 

To put differently, we are looking for a new set of generators $\l\{\bar P_a\r.$, $\bar J_{ab}$, $\bar D$ and $\l.\bar K_a\r\}$,\,\footnote{We should stress that a bar over a generator does not stand for Hermitian conjugation.} whose commutation relations correspond to that of the Euclidean conformal algebra, viz
\be
\label{eq:ComRelEuclidean}
\begin{aligned}
\l [\bar D,\bar P _ a \r ] & =  - i \bar P _ a \ ,  \\
\l [\bar  J _ {a b}, \bar P _ c \r ] & =  i \l ( \d _ {a c} \bar P _ b -\d _ {b c} \bar P _ a \r ) \ ,  \\
\l [\bar K _ a, \bar P _ b \r ] & =  2 i \l ( \d _ {a b} \bar D - \bar J _ {a b} \r ) \ ,  \\ 
\l [ \bar D, \bar K _ a \r ] & =  i \bar K _ a\ , \\
\l [\bar  J _ {a b}, \bar  J _ {c d} \r ] & =  i \l (\d _ {a c} \bar J _ {b d}  + \d _ {b d} \bar J _ {a c}  -  \d _ {b c} \bar J _ {a d} - \d _ {a d} \bar  J _ {b c} \r ) \ ,  \\ 
\l [\bar J _ {a b}, \bar K _ c \r ] & =   i \l ( \d _ {a c} \bar K _ b -\d _ {b c} \bar K _ a \r ),
\end{aligned}
\ee
and whose action on primary (Euclidean) fields $\bar \phi(z)$ is given by
\bea
\label{eq:GeneratorsFieldsEuclideanAction}
\l [ \bar P _ a, \bar \phi (z) \r ] & = & - i \p _ a \bar \phi (z) \ ,
\nn \\ 
\l [ \bar D, \bar \phi (z) \r ] & = & - i \l ( \Delta + z ^a \p _ a \r ) \bar \phi (z) \ ,  \\
\l [ \bar J_ {ab}, \bar \phi (z) \r ] & = & i \l(  \Sigma _ {ab} + z _ a \p _ b - z _ b \p _ a \r ) \bar \phi (z) \ , \nn \\
\l [ \bar K _ a, \bar \phi (z) \r ] & = & i \l ( 2 z _ a z^ b - \d _ a^b z ^ 2 \r ) \p _b \bar \phi (z) + 2 i \l (z _ a \Delta +z ^ b \Sigma _ {a b} \r ) \bar \phi (z) \ , \nn
\eea
with the matrices $\Sigma_{ab}$ now corresponding to representations of $SO(d)$\,\footnote{For the vector representation
\be
\label{eq:vec_rep_spin}
\Sigma_{ab,cd} = \d_{ac}\d_{bd}-\d_{ad}\d_{bc} \ .
\ee
}
\be
[\Sigma_{ab},\Sigma_{cd} ] = \d _{b c}\Sigma_{a d}+\d_{a d} \Sigma_{b
  c}-\d_{a c}\Sigma_{b d}-\d_{b d}\Sigma_{a c} \ .
\ee

Note that the automorphism we are after cannot simply correspond to a Wick rotation. It should be followed by an additional rotation in the $(0,d)$--plane. Namely, it is clear that the generators defined according to
\be
\label{eq:EMSOGenerators}
\widetilde M _ {AB} = i ^ {\l(\d_{A0}+\d_{B0}\r)} M_{AB} \ ,
\ee
have commutation relations identical to that of $M_{AB}$, i.e. 
\be
\label{eucl_conf_cr}
\l [ \widetilde M_ {AB}, \widetilde M _ {CD} \r ] = i \l ( \widetilde M _ {AD} \xoverline g _ {BC} + \widetilde M _
{BC} \xoverline g _ {AD} - 
\widetilde M _ {BD} \xoverline g _ {AC} - \widetilde M _ {AC} \xoverline g _ {BD} \r ) \ ,
\ee
but with a different metric
\be
\xoverline \eta _ {AB} = \mathrm{diag}(-,-,\dots,-,+) \ ,
\label{eucl_metr}
\ee
and, at the same time, modified behavior under Hermitian conjugation    
\be
\label{eucl_unitarity}
\widetilde M _ {AB}^\dagger = \widetilde M_{AB} \ , ~\text{for}~ A,B\neq 0\ ,~~~\text{and}~~~\widetilde M_{0B}^\dagger= -\widetilde M_{0B} \ .
\ee
In other words, the generators $\widetilde M_{AB}$ form a non-unitary representation of the Euclidean conformal group $SO(d+1,1)$. Performing a $\pi/2$ rotation in the $(0,d)$ plane, which is achieved by
\be
\label{eq:Mink_Eucl}
\xoverline M_{AB} = e^{-i\f{\pi}{2}\widetilde M_{0d}}\widetilde M_{AB} e^{i\f{\pi}{2}\widetilde M_{0d}} \ ,
\ee
and using a relation analogous to~\REF{eq:MvsPKDJ} to define\,\footnote{For simplicity, we introduced generators carrying index $d$ rather than $0$.}
\bea
\label{eucl_lor_conf_rel_1}
&&
\xoverline M _ {ij} = \bar  J_{ij}\ ,~~~\xoverline M _ {0i} = \bar J_{di} \ ,~~~\xoverline M _  {d,0} =\f {1} {2} \l(\bar P _ d - \bar K _ d \r ) \ , 
~~~\xoverline M _  {d,i} =\f {1} {2} \l(\bar P _ i - \bar K _ i \r ) \ , \nn \\
&&
\xoverline M_ {d+1,d} = \bar D \ ,~~~\xoverline M _  {d+1,i} =\f {1} {2} \l(\bar P _ i + \bar K _ i \r )\ ,~~~\xoverline M _  {d+1,0} =\f {1} {2} \l(\bar P _ d + \bar K _ d \r ) \ ,
\eea
we get the following map between the Minkowski and Euclidean conformal generators
\bea
\label{eucl_mink_gen_rel}
\bar D = -  \f {i} {2} \l ( P _ 0 + K _ 0 \r )\ ,~~~\bar J _ {ij} = J _
{ij}\ ,~~~\bar J _ {d,i} = \f {1}{2} \l( P _ i - K _ i \r) \ , \nn \\
\bar P _ i = \f {1} {2} \l (P _ i + K _ i \r ) - i J _ {0 i} \ ,~~~
\bar K _ i = \f {1} {2} \l (P _ i +  K _ i \r ) + i J _ {0 i} \ ,  \\
\bar P _ d = D + \f {i} {2} \l ( P _ 0 - K _ 0 \r )\ ,~~~ 
\bar K _ d = D - \f {i} {2} \l ( P _ 0 - K _ 0 \r ) \ , \nn
\eea
with $i,j=1,\ldots, d-1$. It is easy to verify that as anticipated (see Appendix~\ref{app:automorphism} for an alternative automorphism in $d=3$)
\be
\label{eq:autom_main}
\bar J_{ab} = M_{ab}\ ,~~~\bar D = -i M_{d+1,0}\ ,~~~\bar P_a = M^+_a,~~~\bar K_a = M^-_a \ .
\ee
These newly defined generators have the following properties under Hermitian conjugation
\be
\bar J_{ab}^\dagger = \bar J_{ab} \ ,~~~\bar D^\dagger = -\bar D \ ,~~~
\bar P_a^\dagger = \bar K_a \ ,~~~\bar K_a^\dagger = \bar P_a \ ,
\label{eq:HermCFT}
\ee
meaning that such an action defines a non-unitary representation of $SO(1,d+1)$. At the same time, the fields $\bar \phi(0)$ furnish a unitary representation of the Minkowski conformal algebra $SO(2,d)$; in particular, the scaling dimensions $\D$ are in one-to-one correspondence with the spectrum of the ``conformal Hamiltonian'' $M_{d+1,0}$. Every field should be viewed as an element of the Hilbert space, or to put differently, we have the  correspondence 
\be
\bar \phi (0) \leftrightarrow  | \phi \ra\ .
\ee
We note that the scalar product can be defined by specifying it for primary operators\,\footnote{Defined this way, the scalar product is positive definite provided all primary operators satisfy unitarity bounds.}
\be
\la \phi_\a | \phi_\b \ra =\d_{\a\b} \ , ~~~ \la \phi_\a | P_a \phi_\b \ra = 0 \ .
\ee
The states corresponding to $\bar \phi(z)$ can be naturally defined as
\be
\label{eq:state_op_map_CFT_1}
| \phi (z) \ra = e^{iPz} | \phi \ra \ .
\ee

The form of the Operator Product Expansion, which is convergent in CFT~\cite{Luscher:1975js,Mack:1976pa} (see also~\cite{Pappadopulo:2012jk}), is heavily restricted by the conformal symmetry; as an example, for two primary operators it reads
\be
\bar \phi_2(z) \bar \phi_1(0) = \sum_\D C_{\phi_2\phi_1\phi_\D} (z,\p) \bar \phi _\D(0) \ ,
\label{eq:OPEPrimaryCFT}
\ee
 with the sum running over primary fields only and at the same time
 the functions $C_{\phi_2\phi_1\phi_\D}(z, \p)$ are fixed up to
 several structure constants (for more see Appendix~\ref{app:OPE}).\footnote{The OPE in the context of CFTs is very useful when it comes to computing correlation functions. Repeated  use of~(\ref{eq:OPEPrimaryCFT}), breaks down any $n$-point function  to a sum that depends only on the CFT data.}

The OPE endows the space of all fields with an operator algebra, allowing us to define the action of the  operators $\bar \phi(z)$ on the Hilbert space.\footnote{This is reminiscent of how the adjoint representations of Lie algebras are defined.} Namely, the expression~\REF{eq:OPEPrimaryCFT} can be also understood  as
\be
\bar \phi (z) | \phi_1 \ra = \sum _{\D}C_{\phi\phi_1\phi_\D} (z,iP) | \phi _\D \ra \ .
\ee
This construction also enables us to view the states $|\phi \ra$ as obtained by acting with the corresponding field $\bar \phi$ on the vacuum $|0\ra$ (which in turn corresponds to the identity operator), i.e. 
\be
\label{eq:state_op_map_CFT_2}
|\phi\ra = \lim_{z\to 0} \bar \phi(z) | 0 \ra \ . 
\ee

\subsection{Hermitian conjugation}

We should also demonstrate how the Hermitian conjugation of the primary operators constructed this way works. It is obvious from~\REF{eq:GeneratorsFieldsEuclideanAction} that even for real scalar fields $\bar \phi^\dagger(z) \neq \bar \phi(z)$. Instead, in order to preserve the action of the conformal algebra on the  fields,\footnote{See  Appendix~\ref{app:herm_conj}.} we require that scalars transform as
\be
\bar \phi^\dagger(z) = z^{-2\D}\bar \phi\l(I z \r) \ ,
\label{eq:EScalarConjugation}
\ee
vectors as
\be
\bar \phi^\dagger_a(z) = z^{-2\D} I_a^b(z) \bar \phi_b (I z) \ ,
\label{eq:eq:EVectorConjugation}
\ee
 and  rank-$l$ tensors as
\be
\bar \phi^\dagger_{a_1\ldots a_l}(z) = z^{-2\D} I_{a_1}^{b_1}(z)\ldots I_{a_l}^{b_l}(z)\bar \phi_{b_1\ldots b_l} (I z) \ . 
\label{eq:eq:ETensorConjugation}
\ee
The matrix appearing in the above reads\,\footnote{It is easy to check that 
\be
I_{a}^c(z)I_{cb}(z) = I_{ab}(z) \ .
\ee
}
\be
 I_{ab}(z) = \l ( \d _ {a b} - \f{2z_a z_b}{z^2} \r ) \ ,
\ee
and inversion acts on the coordinates as
\be
(Iz)_a = \f{z_a}{z^2} \ .
\ee

\subsection{Explicit coordinate transformations}

Since~(\ref{eucl_mink_gen_rel}) is an automorphism of the conformal algebra, c.f.~\REF{eq:Mink_Eucl}, it is clear that (modulo the analytic continuation) it should be given by a conformal transformation of the coordinates  
\be
x ^ \m \to z^a \ ,
\ee
which can be found in different ways, for instance, using embedding coordinates. However, bearing in mind that later we will turn to NRCFTs, where we cannot avail ourselves of embedding coordinates, it is instructive to find the transformation by comparing the explicit representations of the conformal generators in terms of differential operators in different reference frames. Let us illustrate how that works with an explicit example.

\subsubsection{One-dimensional ``spacetime'' \label{sec:1+0CFT}}

The embedding coordinates in this case~\cite{Salam:1969bwb} correspond to the coordinates $(\xi^0,\xi^1,\xi^2)$ in $\mathbb R ^{2,1}$ (where the action of $SO(2,1)$ is naturally defined) constrained to a cone
\be
(\xi^0)^2-(\xi^1)^2+(\xi^2)^2=0 \ .
\ee
Their relation to the coordinate $x$ parametrizing the initial  one-dimensional ``spacetime'' is given by
\be
x = \f{\xi^0}{\x^2+\x^1} \ .
\ee
Performing a Wick rotation (see~\REF{eq:EMSOGenerators}), followed by another $\pi/2$ rotation in the $(0,1)$--plane (see~\REF{eq:Mink_Eucl}), we get
\be
\l(
\x^0,
\x^1,
\x^2\r)\to\l(
\bar \x^0,
\bar \x^1 ,
\bar \x^2
\r)=
\l(
-\x^1 ,
i \x^0 ,
\x^2 \r) \ ,
\ee
translating into
\be
x\to z= \f{\bar \xi^0}{\bar \x^2+\bar \x^1} = \f{i x+1} {i x-1} \ .
\label{eq:1Dxtoz1}
\ee

We will now derive the same result but in a different way, which can be employed in NRCFTs too.  In~(\ref{eucl_mink_gen_rel}), we found the relation between the generators of the Minkowski and Euclidean conformal algebras. Particularizing to the situation under consideration here, we obtain  
\bea
\label{eq:1dRel}
\bar D & = & - \f {i} {2} \l ( P + K \r )\ , \nn \\
\bar P & = & D + \f {i} {2} \l ( P - K \r ) \ , \\
\bar K & = & D - \f {i} {2} \l ( P - K \r ) \ . \nn
\eea
Expressing the above as differential operators, acting on the functions $f(x)$ and $\bar f(z)$, namely
\be
P = i \p _ x\ ,~~~D= i x \p _ x \ ,~~~K = i x^2 \p _ x \ ,
\ee
and
\be
\bar P = i \p _ z\ ,~~~ \bar D = i z \p _ z\ ,~~~ \bar K = -i z^2 \p _ z \ ,
\ee
we rewrite~(\ref{eq:1dRel}) as 
\bea
z \p _z & = & -\f{i}{2}(1+x^2) \p _x \ , \nn \\
\p _z & = & \f{i}{2}(1-ix)^2 \p _x \ , \\
z^2\p _z & = & \f{i}{2}(1+ix)^2 \p _x \ . \nn
\eea
This leads to the following relation between $x$ and $z$
\be
z= \f{i x+1} {i x-1} \ ,
\label{eq:1Dxtoz2}
\ee
which is identical to~\REF{eq:1Dxtoz1}, as it should.  

The action of the two sets of generators on fields is consistent, provided the following identification between the Minkowski and Euclidean fields is made\,\footnote{As an example, for translations we find
\bea
\l [\bar P, \bar \phi (z) \r ] & = & (z-1)^{-2\D}  \l [D + \f {i} {2}
\l ( P - K \r ), \phi \l ( x(z)\r ) \r ] \nn  \\
&&
=  -i \l(\f{2}{ix-1}\r)^{-2\D}  \l [ \D \l (1-ix \r ) \phi(x)+\f{i}{2} \l ( 1-ix \r )^2 \p \phi (x) \r ] =- i \p \bar \phi (z) \ . 
\eea}
\be
\bar \phi(z)=(z-1)^{-2\D} \phi \l ( i \f{1+z} {1 -z }\r ) \ .
\label{eq:1dEMfields}
\ee
It follows from the above that 
\be
\bar \phi^\dagger (z) = (z-1)^{-2\D}\phi \l ( -i \f{1+z} {1 -z }\r ) = z^{-2\D} \l(\f{1}{z} -1\r)^{-2\D} \phi \l ( i \f{1+\f{1}{z}} {1 -\f{1}{z} }\r )=
z^{-2\D} \phi \l( \f{1}{z}\r ) \ , 
\label{eq:EScalarConjugation1D}
\ee
which is precisely~\REF{eq:EScalarConjugation}.

\subsubsection{Generalization to higher dimensions}

\begin{figure}[H]
\bc
\includegraphics[width=15cm]{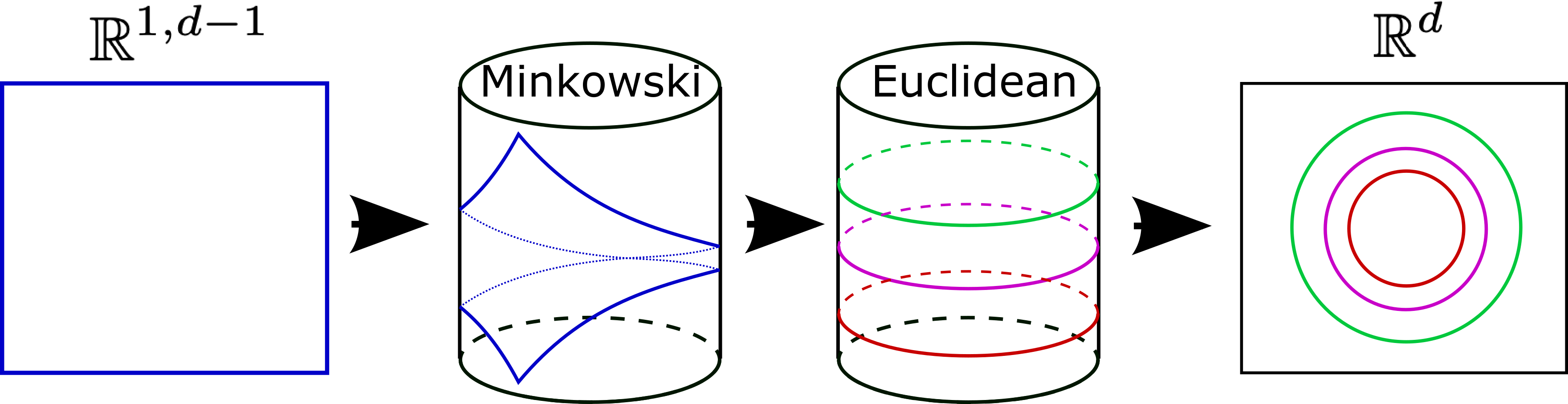}
\ec
\caption{\label{fig:EMCylinder} Minkowski plane to cylinder to Euclidean plane. }
\end{figure}

In general, the coordinate transformation corresponding to the map~\REF{eucl_mink_gen_rel} can be found~\cite{Aharony:1999ti} from~\REF{eq:EMSOGenerators} as follows (see Fig.~\ref{fig:EMCylinder}). First, the Minkowski plane is mapped to the 
$\mathbb R\times \mathbb S^{d-1}$ cylinder by the following change of coordinates (see e.g.~\cite{Hawking:1973uf})\,\footnote{To be more precise, this maps the Minkowski plane to a diamond shaped region of the cylinder. Analytic continuation is implied in extending the map to the whole cylinder.}
\be
x^0 = \f{1}{2} \l ( \tan \f{\tau+\t}{2} + \tan \f{\tau-\t}{2} \r )\ ,~~~| \vec x | = \f{1}{2} \l ( \tan \f{\tau+\t}{2} - \tan \f{\tau-\t}{2} \r )\ ,~~~\f{\vec x }{| \vec x |} = \vec n \ ,
\ee
where $\vec n$ is a unit vector parametrizing the points on a $(d-2)$-dimensional sphere. Then, the Euclidean counterpart of the Minkowski cylinder is obtained by Wick rotating, i.e. for $i\tau=\eta$. Finally, the Euclidean cylinder is mapped to the Euclidean plane by introducing
\be
z_a = e^{\eta}(\vec n \sin \t, \cos \t) \ .
\label{eq:EplaneEcylinder}
\ee

For completeness, the line element written in terms of the different coordinates reads
\bea
ds_{\mathbb R ^{1,d-1}}^2 & = & 
\f{1}{4 \cos^2 \f{\tau+\t}{2}  \cos^2 \f{\tau-\t}{2}} \underbrace{\l (
  d\tau^2-d\t^2-\sin^2\t d\vec n^2 \r ) }_{\text{Minkowski cylinder}} \nn \\
& = & 
 - \f{1}{4 \cos^2 \f{\tau+\t}{2}  \cos^2 \f{\tau-\t}{2}} \underbrace{ \l ( d\eta^2+d\t^2+\sin^2\t d\vec n^2\r ) }_{\text{Euclidean cylinder}}  \\
 & = &
 - \f{e^{-2\eta}}{4 \cos^2 \f{\tau+\t}{2}  \cos^2 \f{\tau-\t}{2}}
ds^2_{\mathbb R^d} \ . \nn
\eea

\subsection{Radial quantization}
\label{sec:radial_quant}

We finish the lightning review by commenting on the radial quantization~\REF{eq:EplaneEcylinder}. Assuming that  all the correlators are obtained from a path integral with a conformally invariant action $S_0[\bar \phi]$ it is straightforward to derive the operator-state correspondence~\cite{Simmons-Duffin:2016gjk,Rychkov:2016iqz}. It is usually the case that a conformally invariant, unitary, theory can be put on a curved manifold in a Weyl invariant way (for examples of non-unitary theories defying this assumption see~\cite{Karananas:2015ioa}). In other words, the action $S_0[\bar \phi]$ capturing the dynamics of a CFT can be generalized to $S[\bar \phi, g_{ab}]$ that also depends on the background metric
\be
S_0[\bar \phi]=S[\bar \phi,\d_{ab}] \to S[\bar \phi, g_{ab}]  \ ,
\ee
such that
\be
\label{eq:weyl_inv_rel_action}
S[e^{-\D_W \s} \bar \phi, e^{2\s}g_{ab}] = S[\bar \phi, g_{ab}] \ ,
\ee
where $\D_W$ is the Weyl weight of the field $\bar \phi$; if a field with scaling dimension $\D$ has $n_L$ lower and $n_U$ upper indices, then its Weyl weight is given by
\be
\D_W = \D + n_U-n_L \ .
\ee
Considering a diffeomorphism corresponding to the coordinate transformation~\REF{eq:EplaneEcylinder}, i.e. 
\bea
S _ 0 \l [ \bar \phi ^ {a\dots}  _ {b \dots} (z) \r ]  = 
S \l [ \bar \phi ^ {'a\dots}  _ {b \dots} (\eta,\vec n), e ^ {2 \eta} (d \eta ^ 2 + d \Omega _{d-1} ^ 2) \r ] \ ,
\eea
followed by a Weyl rescaling with $\s = \eta$, leads to 
\bea
S \l [ \bar \phi ^ {'a\dots}  _ {b \dots} (\eta,\vec n), e ^ {2 \eta} (d \eta ^ 2 + d \Omega _{d-1} ^ 2) \r ] =  S \l [ e^{\D_W \eta} \bar \phi^ {'a\dots}  _ {b \dots} (\eta,\vec n), d \eta ^ 2 + d \Omega _{d-1} ^ 2 \r ] \ .
\eea
The fact that the theory is Weyl invariant, c.f.~(\ref{eq:weyl_inv_rel_action}), translates into
\bea 
S \l [ e^{\D_W \eta} \bar \phi^ {'a\dots}  _ {b \dots} (\eta,\vec n), d \eta ^ 2 + d \Omega _{d-1} ^ 2 \r ] = S \l [ \hat {\phi} ^ {a\dots}  _ {b \dots} (\eta, \vec n), d \eta ^ 2 + d \Omega _{d-1} ^ 2 \r ] \ ,
\eea
implying of course the  equivalence of the system on the plane and on the cylinder, i.e.
\bea
S _ 0 \l [ \bar \phi ^ {a\dots}  _ {b \dots} (z) \r ]  = 
S \l [ \hat {\phi} ^ {a\dots}  _ {b \dots} (\eta, \vec n), d \eta ^ 2 + d \Omega _{d-1} ^ 2 \r ] \ ,
\eea
provided we define 
\be
\hat {\phi} ^ {a\dots}  _ {b \dots} (\eta, \vec n) \equiv e^{\D_W \tau} \f{\p z^{'a}}{\p z^c} \dots \f{\p z^b}{\p z^{'d}} \dots \bar \phi^ {c\dots}  _ {d \dots} (z) \ ,
\ee
and $z'$ stands for the cylinder coordinates  parametrized by $\eta$ and $\vec n$.

It follows from \REF{eq:EplaneEcylinder} that the generator controlling translations in the cylinder's ``time'' $\eta$ is identical to dilatations $\bar D$ on the Euclidean plane. Indeed, on the cylinder we have
\bea
\label{eq:dilat_barred_cyl}
\l[ \bar D,\hat {\phi} ^ {a\dots}  _ {b \dots} (\eta, \vec n) \r ] & = &
e^{\D_W \eta} \f{\p z^{'a}}{\p z^c} \dots \f{\p z^b}{\p z^{'d}} \dots
\l[ \bar D,  \bar \phi^ {c\dots}  _ {d \dots} (z) \r ] \nn \\ 
& = &  - i e^{\D_W \eta} \f{\p z^{'a}}{\p z^c} \dots \f{\p z^b}{\p z^{'d}} \dots (\D_W + \p _ \eta) \bar \phi^ {c\dots}  _ {d \dots} (z) = 
-i \p _ \eta \hat {\phi} ^ {a\dots}  _ {b \dots} (\eta, \vec n) \ ,
\eea
which means that the eigenstates of the Hamiltonian (controlling the evolution in real time) on the cylinder are in one-to-one with the dimensions of the fields. There are other manifolds where Hamiltonians are related to different generators of the Euclidean conformal algebra. For instance, in~\cite{Alday:2007mf} it was shown that for $AdS_{d-1}\times\mathbb S^1$, the eigenstates of the Hamiltonian are given by the twists of the corresponding operators.

\section{Non-Relativistic CFTs}
\label{sec:NRCFT}

In this section we present a similar construction for non-relativistic conformal field theories. 

\subsection{The Schr\"odinger algebra}

The Schr\"odinger algebra comprises the generators of temporal and spatial translations $H$ and $P_i$, respectively, spatial rotations $J_{ij}$, Galilean boosts $K_i$, dilatations $D$ and special conformal transformations $S$. As usual, the algebra is also centrally extended by adding the particle number operator $Q$ commuting with the rest of the generators. 

The standard commutation relations for the Galilei algebra 
\bea
\l [ J _ {i j}, J _ {k l} \r ] & = & i \l ( J _ {j l} \d _ {i k} + J _
{i k} \d _ {j l} - J _ {i l} \d _ {j k} - J _ {j k} \d _ {i l}  \r ) \
, \nn \\
\l [ J _ {i j}, P _ {k} \r ] & = & i \l ( \d _ {i k} P _ j - \d _ {j
  k} P _ i  \r ) \ , \\
\l [ J _ {i j}, K _ {k} \r ] & = & i \l ( \d _ {i k} K _ j - \d _ {j
  k} K _ i  \r ) \ , \nn
\label{eq:GalileiCR1}
\eea
should be supplemented by the way the momenta and Hamiltonian are transformed under boosts, 
\be
\l [ K _ {i}, P _ {j} \r ] = - i \d _ {i j} Q \ , ~~~\l [ K _ {i}, H
\r ] =- i P _ i\ , \\
\label{eq:GalileiCR2}
\ee
as well as the nonvanishing commutators involving dilation and special conformal transformations
\be
\ba{lll}
\l [ D, H \r ] = - 2 i H \ , &  \l [ D, P _ i \r ] = - i P _ i \ , \\
\l [ D, K _ i \r ] = i K _ i \ , & \l [ D, S \r ] = 2 i S \ , \\
\l [ S, H \r ] = i D \ , &\l [ S, P _ i \r ] = i K _ i \ .
\ea
\label{eq:SchrodingerCR}
\ee
Several comments are in order here. From~(\ref{eq:GalileiCR1}), we notice that $P_i$ and $K_i$ (we are in $\mathbb R \times \mathbb R^d$, and the indices $i,j,\dots$ run from $1$ to $d-1$) are vectors of $SO(d-1)$. Meanwhile, inspection of~(\ref{eq:SchrodingerCR})  
reveal that the scaling dimensions of $H$, $P_i$, $K_i$ and $S$ are $2$, $1$, $-1$ and $-2$, respectively.

\subsection{Representations \label{sec:Representations}}

The representations of the Schr\"odinger algebra can be build in the following way. We can choose the subalgebra spanned by $Q$, $E\equiv H-S$ and $J_{ij}$
to label any state with the eigenvalues $m$, $\eps$, $j$ and $\s$ of $Q$, $E$ and the spins corresponding to $SO(d-1)$. For instance, in $d=4$  we have
\bea
E |m, \eps, j, \s \ra & = & \eps | m, \eps, j, \s \ra \ , \nn \\
J_{ij}J_{ij} |m, \eps, j, \s \ra & = & j(j+1) |m, \eps, j, \s \ra \ , \\
J_3 |m, \eps, j, \s \ra & = & \s |m, \eps, j, \s \ra \ , \nn \\
Q |m, \eps, j, \s \ra & = & m |m, \eps, j, \s \ra\ . \nn
\eea
It is straightforward to show that for the following linear combinations
\be
\label{eq:SchRepFG}
F_i^\pm \equiv \f{K_i\pm i P_i}{\sqrt{2}} \ ,~~~G^\pm = \f{1}{2} \l [
D \pm i \l ( H+S \r ) \r ] \ ,
\ee
we have
\be
\l [ E, F^\pm_i \r ] = F^\pm_i \ ,~~~\l [ E, G^\pm \r ] = 2G^\pm \ ,
\ee
therefore these operators act as raising and lowering operators for the eigenvalues of $E$. Noting also that $(G^-, F_i^-)$ and $(G^+, F_i^+)$ form two Abelian subalgebras, we can define a lowest weight state as
\be
F_i^- |m, \eps, j, \s \ra = G^- |m, \eps, j, \s \ra=0 \ ,
\ee
and generate the whole Hilbert space by acting on it repeatedly with $(G^+, F_i^+)$. It may be advantageous to introduce spin eigenstates $F^+_\a$ with $\a = (+,0,-)$
\be
F^+_\pm \equiv \f{1}{\sqrt{2}} \l ( F_{1}\pm F_2 \r ) \ , ~~~F_0^+
\equiv F^+_3 \ ,
\ee
such that
\be
\l [ J_3, F^+_\a \r] = \a F^+_\a \ .
\ee
As a result the Hilbert space is given by a span of vectors of the form
\be
|m, \eps, j, \s; n_+,n_0,n_-,k \ra = \l ( F^+_+ \r )^{n_+} \l ( F^+_0
\r )^{n_0} \l ( F^+_- \r )^{n_-} \l ( G^+ \r )^{k} |m, \eps, j, \s \ra
\ ,
\ee
with $n_+$, $n_0$, $n_-$ and $k$ integers; the corresponding
eigenvalues of $E$ and $J_3$ are
\bea
E |m, \eps, j, \s; n_+,n_0,n_-,k \ra & = & (\eps+n_++n_0+n_-+2k)|m,
\eps, j, \s; n_+,n_0,n_-,k \ra \ , \nn \\
J_3 |m, \eps, j, \s; n_+,n_0,n_-,k \ra & = & (j+n_+-n_-)|m, \eps, j,
\s; n_+,n_0,n_-,k \ra \ .
\eea
We note in passing that the four-dimensional Schr\"odinger algebra  possesses three Casimir operators~\cite{Perroud:1977qh, Andrzejewski:2012nf}. One of them is obviously $Q$,  while the others are  quadratic and quartic in the generators of the algebra.

\subsection{Action of the algebra on fields}

The way the algebra acts on fields is derived in the standard way by constructing a representation induced from that of a subalgebra generated by 
$(Q,J,D,K,S)$---this obviously leaves the origin $(t=0,\vec x =0)$ invariant, therefore, we can define its action on fields there. Namely, since the generators $Q,J,D$ commute with each other,  we have
\bea
\label{NRCFT_irrep_1}
\l [ D, \phi(0) \r] & = & - i \D \phi(0)\ , \nn \\
\l[ Q, \phi(0) \r] & = & -m \phi(0)\ , \\
\l[ J_{ij}, \phi(0) \r] & = & i\Sigma _{ij} \phi(0) \ , \nn
\eea
where $\D, m$ are numbers and $\Sigma_{ij}$ is a finite-dimensional matrix corresponding to an irreducible representation of the $SO(d-1)$ spatial rotations. Assuming that a subset of fields is closed under the action of $K_i$ and $S$, in other words
\bea
\l[ S, \phi(0) \r] = i s \phi(0) \ ,~~~\l[ K_i, \phi(0) \r] =i \kappa
_i \phi(0) \ ,
\eea
with $s$ and $\kappa_i$ matrices, it follows that 
\be
\l [ \D, \kappa _ i \r ] = \kappa_ i \ ,~~~ \l [ \D, s \r ] = 2  s \ ,
\ee
which means that $s=\kappa=0$ and operators $S$ and $K_i$ are realized trivially
\be
\l[ S, \phi(0) \r]  =  0 \ , ~~~ \l[ K_i, \phi(0) \r]  =  0 \ .
\label{NRCFT_irrep_2} 
\ee
Now from
\be
\l [ H, \phi(t,\vec x)\r ] = - i \p _ t \phi (t,\vec x) \ , ~~ \l [
P_i, \phi(t,\vec x)\r ] = i \p _ i \phi (t,\vec x) \ ,
\ee
equivalently,
\be
\phi (t,x_i) = e^{i Ht - i P_i x_i} \phi (0)e^{-i Ht + i P_i x_i} \ , 
\ee
we get\,\footnote{For instance,
\bea
\l[ K_i, \phi(t,\vec x) \r] & = & \l[ K_i, e^{i Ht - i P_i x_i} \phi (0)e^{-i Ht + i P_i x_i} \r] = e^{i Ht - i P_i x_i} \l[ e^{-i Ht + i P_i x_i} K_i e^{i Ht - i P_i x_i}, \phi (0) \r] e^{-i Ht + i P_i x_i} \nn \\
& = & e^{i Ht - i P_i x_i} \l[ K_i +tP_i -x_i Q , \phi (0) \r] e^{-i Ht + i P_i x_i} =
e^{i Ht - i P_i x_i} \l ( i t \p_i \phi (0) + x_i m \phi (0) \r ) e^{-i Ht + i P_i x_i} \nn \\
& = & i \l ( t \p _ i - i m x_i \r ) \phi(t,\vec x) \ , \nn
\eea
where we used Eq.~(\ref{eq:app_transl_action}) from  Appendix~\ref{app:schro_al}.}
\bea
\label{eq:FieldsNRCFT}
\l [ D, \phi(t,\vec x) \r] & = & - i \l ( \D + 2 t \p _t+x_i \p _i \r
) \phi(t,\vec x) \ , \nn \\
\l[ Q, \phi(t,\vec x) \r] & = & -m \phi(t,\vec x) \ , \nn \\ 
\l[ J_{ij}, \phi(t,\vec x) \r] & = &  i \l ( \Sigma _{ij} \phi(x) +
x_i \p _ j - x_j \p _i \r ) \phi (x) \ , \\ 
\l[ S, \phi(t,\vec x) \r] & = & i \l ( t \D + t^ 2 \p _t + x_i t \p _
i - \f {i}{2} m x_i^2 \r ) \phi(t,\vec x) \ , \nn \\ 
\l[ K_i, \phi(t,\vec x) \r] & = & i \l ( t \p _ i - i m x_i \r )
\phi(t,\vec x) \ . \nn
\eea

\subsection{Automorphism}

To be able to construct the Hilbert space from fields the way it was done for the case of relativistic CFTs, we need to find an automorphism (up to an analytic continuation) of the Schr\"odinger generators
\be
H, P_i, J_{ij}, D, K_i, S~~~\to~~~\bar H, \bar P, \bar J_{ij}, \bar D, \bar K_i, \bar S,
\ee
such that the new, barred, ones have the appropriate conjugation
properties. 

The commutation relations involving the new generators, as well as their action on (barred) fields are of course similar to the ones we presented above. From the ``barred counterpart'' of~(\ref{NRCFT_irrep_1}), we notice that $\bar J$ and $\bar Q$ should be Hermitian while $\bar D$ should be anti-Hermitian
\be
\bar J_{ij}^\dagger = J_{ij} \ ,~~~\bar Q^\dagger = \bar Q\ ,~~~\bar D
^\dagger = - \bar D \ .
\ee
As $Q$ is just the central charge we can immediately fix $\bar Q=Q$. To preserve the structure of the commutation relations~(\ref{eq:GalileiCR2}) and~(\ref{eq:SchrodingerCR}), one should impose specific conjugation properties for the rest of the operators. For instance, 
\be
\l [ \bar D, \bar H^\dagger \r ] =-\l [ \bar D^\dagger, \bar H^\dagger
\r ] = \l [ \bar D, \bar H \r ]^\dagger = (- 2 i \bar H)^\dagger = 2 i
\bar H^\dagger \ ,
\ee
which can be satisfied, provided we identify
\be
\bar H^\dagger = \bar S \ .
\label{eq:HSrelation}
\ee
This automatically implies
\be
\l [ \bar S,\bar H \r ]=-\l [ \bar H , \bar S \r ]=-\l [ \bar S^\dagger , \bar H^\dagger  \r ]=\l [ \bar S, \bar H \r ]^\dagger = (i \bar D)^\dagger = i \bar D \ ,
\ee
which is essentially the same as for the relativistic conformal group~(\ref{eq:HermCFT}). The difference comes for $P_i$ and $K_i$. Indeed, we have
\be
\l [ \bar D, \bar P_i^\dagger \r ] =-\l [ \bar D^\dagger, \bar
P_i^\dagger \r ] = \l [ \bar D, \bar P_i \r ]^\dagger = (- i \bar
P_i)^\dagger = i \bar P_i^\dagger \ ,
\ee
which is clearly satisfied for
\be
\bar P_i^\dagger = \alpha_1 \bar K _ i \ ,
\ee
with $\a_1$ a complex number. Similarly,
\be
\l [ \bar D, \bar K_i^\dagger \r ] =-\l [ \bar D^\dagger, \bar
K_i^\dagger \r ] = \l [ \bar D, \bar K_i \r ]^\dagger = ( i \bar
K_i)^\dagger = - i \bar K_i^\dagger \ ,
\ee
for which 
\be
\bar K_i^\dagger = \a_2 \bar P _ i \ ,
\ee
where $\a_2$ is also a complex number. Note that for operators with
$\bar Q \neq 0$, the constants  
$\a_1$ and $\a_2$ are constrained as\,\footnote{For operators with $\bar Q=0$ there is no constraint for $\a_1~\&~\a_2$ except~(\ref{eq:ab_constr1}),  and one may choose
\be
\bar P_i^\dagger = \bar K_i\ ,~~~\bar K_i^\dagger = \bar P_i \ .
\ee} 
\be
\a_1 \a_2 = -1 \ ,
\ee
since
\be
 [\bar K_i,\bar P_j]^\dagger = \a_1\a_2 \l [\bar K_j,\bar P_i \r ] =i\d_{ij}\bar Q  \ ,
\ee
and
\be
 [\bar K_i,\bar P_j]= - i\d_{ij}\bar Q \ .
\ee

Taking also into account that
\be
\bar P_i = (\bar P_i^\dagger)^\dagger\ ,~~~\bar K_i = (\bar K_i^\dagger)^\dagger\ ,
\ee
leads to
\be
\a^*_{1,2}=-\a_{1,2} \ ,
\label{eq:ab_constr1}
\ee
so we choose $\a_1=\a_2=i$, i.e.
\be
\bar P_i^\dagger = i \bar K_i \ ,~~~\bar K_i^\dagger = i \bar P_i \ . 
\label{eq:PKrelation}
\ee

Now we move to finding the needed automorphism. Noting that the subalgebra generated by $H$, $S$ and $D$ is identical (modulo the sign of $S$) to that generated by $P_0$, $K_0$ and $D$, we can guess immediately the appropriate transformation. Namely (compare with \REF{eq:Mink_Eucl}) we consider an automorphism of the Schr\"odinger algebra generated by a linear combination of $H$ and $S$
\be
\exp \l [ \f{\pi} {4} (H+S) \r ] \ .
\label{eq:SchrodingerMink_Eucl}
\ee
Using the results of Appendix \ref{app:NRbarTransform} we obtain ($\bar Q = Q$ and $\bar J _{ij} = J _{ij}$)
\bea
\bar P_i & = & \f{1}{\sqrt{2}} \l( P_i  - i K_i \r ) \ , \nn \\
\bar K_i & = & \f{1}{\sqrt{2}} \l( -i P_i + K_i \r ) \ , \nn \\
\bar S& = & \f{1} {2} \l ( H+S + i D \r ) \ , \\
\bar H & = & \f{1} {2} \l ( H + S  - i D \r ) \ , \nn \\
\bar D & = & - i \l ( H -S \r ) \ . \nn
\label{eq:AutomNRCFT}
\eea
Comparing with \REF{eq:SchRepFG} we see that, as we wanted, the newly constructed generators are in one to one correspondence with those used in Section~\ref{sec:Representations} to construct the representations of the Schr\"odinger algebra
\be
\bar P_i = - i F_i^+ \ ,~~~\bar K_i = F_i^- \ ,~~~\bar S = i G^-\
,~~~\bar H = -iG^+\ ,~~~\bar D = - i E \ .
\ee
 Using expressions completely analogous to the ones presented in Eqs.~\REF{eq:state_op_map_CFT_1}-\REF{eq:state_op_map_CFT_2} for the non-relativistic case finalizes the construction of the Hilbert space in terms of local operators.

\subsection{Coordinate transformations}

We will now identify the coordinate transformation  that actually corresponds to the automorphism of the Schr\"odinger algebra discussed above. Let us denote the new coordinates by $(s,\vec z)$. 
As in Sec.~(\ref{sec:1+0CFT}), we shall use the explicit representation of the generators in the two coordinate systems in terms of differential operators, i.e.
\bea
P_i = -i \p^x _ i, ~H= i \p _t\ ,&&D = i (2 t \p _ t +  x_i \p^x _i)\
,~~~K_i = -i t \p^x _i \ , \nn \\
S= - i (t^2 \p _t +x_i t \p^x_i)\ ,&& J _ {ij} = - i (x_i \p^x _j -
x_j \p^x _i) \ ,
\eea
and
\bea
\bar P_i = - i \p^z _ i\ ,~~~\bar H= i \p _s \ , && \bar D = i (2 s \p
_ s +  z_i \p^z _i) \ ,~~~\bar K_i = -i s \p^z _i \ , \nn \\
\bar S= -i (s^2 \p _t +z_i s \p^z_i) \ , && \bar J _ {ij} = - i (z_i
\p^z _j - z_j \p^z _i) \ .
\eea
From the above expressions we obtain 
\be
\label{eq:NRCFTCoordinateTransform}
i s = \f {1+it}{1-it} \ ,~~~ z_i = \sqrt{2} \f {x_i} {1-it} \ .
\ee

The corresponding transformation of a scalar primary field can be deduced from~(\ref{eq:NRCFTGeneralTransformation}) (see Appendix~\ref{app:NRCFTGeneralTransformation} for the explicit derivation) and reads{\,\footnote{For tensor fields the transformation will involve matrix factors analogous to the ones in Eq.~(\ref{eq:eq:ETensorConjugation}). The general form may be found using the results from Appendix~\ref{app:NRCFTGeneralTransformation}.}
\be
\phi (t, \vec x) = \f{2^{\D/2}} {(1-it)^\D} \exp \l ({\f {m \vec x^2}{2(1-it)}} \r ) \bar \phi \l ( -i \f {1+it} {1-it},  \f {\vec x
  \sqrt{2}} {1-it}\r ) \ .
\label{eq:FieldTransNRCFT}
\ee
Given the relation between fields in different frames we can
find how $\bar \phi$ transforms under Hermitian
conjugation\,\footnote{For a closer analogy with the relativistic case we can consider Euclidean time $\s = i s$ and introduce a new field 
\be
\label{eq:EucledeanNRCFTfield}
\bar \vp (\s,\vec z ) = \bar \phi (-i\s, \vec z) \ ,
\ee 
whose Hermitian conjugation
\be
\bar \vp^\dagger (\s, \vec z) = \s^{-\D} \exp \l ({\f {m \vec z
    ^2}{2\s} \f{1-\s}{1+\s}} \r ) \bar \vp^* \l ( \f{1}{\s},\f{\vec
  z}{\s} \r ) \ ,
\ee
is reminiscent of its relativistic counterparts \REF{eq:EScalarConjugation} and \REF{eq:EScalarConjugation1D}.}
\be
\bar \phi ^\dagger (s, \vec z) = (-is)^{-\D}  \exp \l (i{\f {m \vec z
    ^2}{2s} \f{1+is}{1-is}} \r ) \bar \phi^*\l ( \f{1}{s},\f{i\vec
  z}{s} \r ) \ .
\ee
One can check that the so-defined Hermitian conjugation preserves the
action of the Schr\"odinger algebra on fields~(\ref{eq:FieldsNRCFT})
and at the same time is consistent with the conjugation properties~(\ref{eq:HSrelation}) and~(\ref{eq:PKrelation}). 

\subsection{NRCFT and geometric data}

It is well-known~\cite{Son:2005rv,Son:2013rqa,Son:2008ye,Brauner:2014jaa,Karananas:2016hrm} that coupling a non-relativistic system to a non-trivial gravitational
background (geometry) can be achieved by introducing appropriate gauge
fields (analogues of metric and connection in general relativity, see
Appendix~\ref{app:NewtonCartan}). Namely, the needed geometric data
are the temporal and spatial parts of a vielbein---$n_\m$, $e^i_\m$,
respectively---and a gauge field $A_\m$ corresponding to the $U(1)$
transformations generated by the particle number operator. 

The non-relativistic conformal transformations can be defined similarly to the relativistic ones. Starting form a trivial (flat) background, corresponding to
\be
n_\m(t,\vec x )= \d_\m ^0 \ ,~~~e_\m ^i (t,\vec x )= \d_\m^i\ ,~~~
A_\m(t,\vec x ) =0 \ ,
\ee
we call conformal those coordinate transformations that lead to a change of the geometric data by a conformal factor
\bea
\label{eq:Tvielbein}
n'_\m(t', \vec x') & \simeq & f^2(t,\vec x) \d_\m ^0  \ , \\
\label{eq:Xvielbein}
e^{'i}_\m(t', \vec x') & \simeq & f(t,\vec x) \d_\m ^i \ , \\
\label{eq:Ageom}
A'_\m(t', \vec x') & \simeq & 0 \ .
\eea
Here ``$\simeq$'' is understood as an equality modulo possible gauge
transformations listed in
Table~\ref{tab:TransformationsNewtonCartan}. Introducing the infinitesimal change of coordinates
\be
t' = t+\x_t \ , ~~ x_i' = x_i +\x_i \ ,
\ee
and denoting $f = 1+\psi$, we see from \REF{eq:Tvielbein} that $\x_t=\x_t(t)$ and
\be
\label{eq:psi_t}
\psi(t) = -\f{1}{2}\p_t \x_t \ .
\ee
The temporal ($\m=t$) component of \REF{eq:Xvielbein} can be satisfied
by using a boost transformation with parameter $v_i = -\p_t
\x_i$. For the spatial ($\m=j$) components of \REF{eq:Xvielbein} we get
\be
(1+\psi)\d_{ij} = \d_{ij}-\p_j\x_i +r_{ij} \ ,
\ee
where $r_{ij}$ is an antisymmetric matrix corresponding to gauge rotations of the vielbein (see Table~\ref{tab:TransformationsNewtonCartan}). 
Symmetrizing the above  and using \REF{eq:psi_t} we obtain 
\be
\p_i \x_j + \p_j \x_i = \d_{ij} \p_t \x_t \ ,
\ee
whose solution is clearly at most linear in $x_i$
\be
\x_i = \f{1}{2} \p_t \x_t x_i + b_{ij} (t) x_j + c_i(t) \ ,~~~
b_{ij}(t)= - b_{ji}(t) \ .
\ee
Bearing in mind that the $U(1)$ gauge field produced by the boost transformation with parameter $v_i = -\p_t \x_i$
\be
A_t = 0 \ , ~~~A_i = -\p_t \x_i = -\l(\f{1}{2} \p^2_t \x_t x_i + \p_t
b_{ij}(t)x_j+\p_t c_i(t) \r)\ ,
\ee
can be eliminated by a $U(1)$ transformation only if $\p^2_t \x_t =\const$, $\p_t b_{ij}(t) = 0$, and $\p_t c_i(t)=\const$, we find
the following expression for the infinitesimal transformations
\be
\x_t = c_t + 2 \lambda t + \m t^2 \ , ~~ \x_i = c_i +d_i t
+b_{ij}x_j+\lambda x_i + \m t x_i \ .
\ee
The above correspond to time and space translations, rotations, boosts, dilations and special conformal transformations.

\subsection{The analogue of radial quantization}

In order to have complete analogy with the relativistic case,  we will show here what happens if one considers dilatations generated by $\bar D$ as the corresponding Hamiltonian. It was shown in~\cite{Karananas:2016hrm} that with minor assumptions an NRCFT can be coupled to a non-trivial background (geometric data) in a Weyl invariant manner, i.e. one constructs the curved-spacetime counterpart of a Schr\"odinger-symmetric action
\be
S_0[\phi] = S [\phi, 0, \d_{\m}^0,\d_{\m}^i] \to S [\phi, A_\m, n_\m, e_\m^i] \ ,
\ee
such that it is manifestly invariant under Weyl rescalings
\be
S [\phi \Omega^{-\D}, A_\m, n_\m \Omega^{2\D}, e_\m^i \Omega^{\D}]=S [\phi, A_\m, n_\m, e_\m^i] \ ,
\ee
with $\Omega$ the conformal factor. 

For the case at hand we start from flat space and consider the following change of coordinates\,\footnote{Note that from~\REF{eq:NRCFTCoordinateTransform} and~(\ref{eq:harm_1}), we get 
\be
t= \tan \tau \ ,~~~x_i = \f{y_i}{\cos \tau} \ .
\ee
These coordinates parametrize the so-called harmonic or oscillator frame. }
\be
\label{eq:harm_1}
i s = e^{2i \tau} \ , ~~~z_i = \sqrt 2 y_i e^{i\tau}\ ,
\ee
which amounts to replacing the temporal and spatial Kronecker symbols by the corresponding vielbeins (the relevant transformation properties for the geometrical quantities can be derived using the Table~\ref{tab:TransformationsNewtonCartan}) 
\be
n_\m = 2e^{2i\tau} (1,\vec 0),~~ e^i_\m = \sqrt 2 e^{i\tau} \l (i\vec y , \mathbb{1}\r ) \ .
\ee
We find that the action becomes 
\be
S\l[\bar \phi(s,\vec z),0,\d_\m^s,\d_\m^i\r]=S\l[\bar \phi'(\tau,\vec
y),0,n_\m,e_\m^i\r] \ , 
\ee
with $\bar \phi'(\tau, \vec y) = \bar \phi(s,\vec z)$. Then, we perform a Weyl rescaling with $\Omega=\sqrt{2} e^{i\tau}$, yielding 
\be
S\l[\bar \phi(s,\vec z),0,n_\m,e_\m^i\r] = S\l[2^{\D/2}e^{i\D\tau}\bar
\phi'(\tau,\vec y) ,0,\d_\m^\tau ,\l (i\vec y , \mathbb{1}\r )\r] \ .
\ee
Next, in order to bring the vielbein to its original form, we consider a boost with parameter $v_i=-iy_i$, such that
\be
S\l[2^{\D/2}e^{i\D\tau}\bar \phi'(\tau,\vec y) ,0,\d_\m^\tau ,\l
(i\vec y , \mathbb{1}\r )\r] =S\l[2^{\D/2}e^{i\D\tau}\bar
\phi'(\tau,\vec y) ,A_\m,\d_\m^\tau ,\d_\m^i\r] \ ,
\ee
which results also in the generation of the following $U(1)$ gauge field
\be
A_s = \f{\vec y^2}{2} \ , ~~~A_i = - i y_i \ .
\ee
Lastly, performing a $U(1)$ gauge transformation with parameter $\a =
i \f {\vec y^2}{2}$ we can eliminate the spatial part $A_i$, to obtain 
\be
S_0[\bar \phi(s, \vec z)] = S\l[\hat \phi (\tau, \vec y) \ ,\hat
A_\m(\tau,\vec y),\d_\m^\tau, \d_\m^i \r] \ ,
\label{eq:FramesRelation}
\ee
with
\be
\hat \phi (\tau, \vec y) = 2^{\f{\D}{2}} e^{i \D \tau} e^{\f{m \vec
    y^2}{2}} \bar \phi(s,\vec z)\ , ~~~ \hat A_\tau = \f{\vec y^2}{2} \ .
\ee
Note that the form of the transformed field $\hat \phi$ is consistent with \REF{eq:NRCFTGeneralTransformation}. 
Equation \REF{eq:FramesRelation} is the analogue of putting a CFT on
the cylinder. It tells us that the systems with and without harmonic potential are equivalent.

A straightforward computation -- completely analogous to the one we explicitly carried out in Sec.~\ref{sec:radial_quant}, see Eq.~(\ref{eq:dilat_barred_cyl}) -- reveals that in this frame, dilatation $\bar D$ acts on operators $\hat \phi$ as time translations, i.d. it plays the role of a Hamiltonian
\bea 
\l[i \bar D, \hat \phi(\tau,\vec y) \r] &=&2^{\f{\D}{2}} e^{i \D \tau}
e^{\f{m \vec y^2}{2}}\l[i \bar D, \bar \phi (s,\vec z) \r]\nn\\
&=&\, 2^{\f{\D}{2}} e^{i \D \tau}
e^{\f{m \vec y^2}{2}}\l(\D+2s\p_s+z_i \p_i^z\r) \bar \phi (s,\vec
z) =-i\p_\tau \hat \phi(\tau,\vec y) \ . 
\eea
Clearly, this means that the spectrum of the Hamiltonian in the
harmonic frame is in one to one correspondence with the spectrum of
scaling dimensions of operators.\footnote{Similarly to
  \REF{eq:EucledeanNRCFTfield}, introducing the Euclidean version of the field
\be
\hat \vp (\eta,\vec y ) = \hat \phi (-i\eta, \vec y) \ , ~~~\eta = i \tau
\ ,
\ee 
we get
\be
\l[\bar D, \hat \vp(\eta,\vec y) \r] = -i\p_\eta \hat \vp(\eta,\vec y)
\ ,
\ee
which is identical to \REF{eq:dilat_barred_cyl}.}

\section{Conclusions}
\label{sec:conclusions}

A powerful tool when it comes to studying relativistic and nonrelativistic conformal field theories is the operator-state map, in particular, the correspondence between the scaling dimensions of operators and the ``energy spectrum'' of the associated states. 
 
In this paper we introduced an algebraic in nature perspective on the aforementioned correspondence. The crucial observation is that the Hilbert space associated with the conformal algebra may be constructed by Euclidean fields. This implies that the operator-state map is obtained by establishing the appropriate relation (automorphism) between the generators of the Minkowski-space conformal algebra and their Euclidean-space counterparts together with the OPE. 
 
Using the derivation in CFT as a guide, we extended the construction to NRCFTs, for which we recover the well-known correspondence between the operators in the theory and states of the system supplemented by an oscillator potential.

\section*{Acknowledgements}

The work of AM was partially supported by the Swiss National Science Foundation under contract 200020-169696.

\appendices

\section{An alternative automorphism in $d=3$ dimensions \label{app:automorphism}}

In this appendix we discuss an alternative to the automorphism discussed in the main text. For clarity, we confine ourselves to $d=3$. As before, we denote with $M_{AB}$ the generators of $SO(2,3)$ that satisfy the commutation relations~(\ref{mink_conf_cr}), i.e.
\be
\l [ M _ {AB}, M _ {CD} \r ] = i \l ( M _ {AD} \eta _ {BC} + M _ {BC}
\eta _ {AD} - M _ {BD} \eta _ {AC} - M _ {AC} \eta _ {BD} \r ) \ ,
\ee
with the five-dimensional metric
\be
\eta_{AB} = \l ( +, -,-,-, + \r ) \ .
\ee
The automorphism we are after corresponds to introducing the rotated, ``barred generators'' $\bar M_{AB}$, which are related to the original ones via successive $\pi/2$ rotations in the $(0,3)$-, $(4,1)$-, and $(0,2)$- planes:
\be
\bar M_{AB} = e^{-i\f{\pi}{2}M_{02}}e^{i\f{\pi}{2}M_{41}}e^{i\f{\pi}{2}M_{03}}M_{AB} e^{-i\f{\pi}{2}M_{03}}e^{-i\f{\pi}{2}M_{41}}e^{i\f{\pi}{2}M_{02}} \ .
\ee
Explicitly, the above yields
\bea
&&
\bar M_{01} = M_{43} \ , ~~~ \bar M_{02} = M_{30} \ , ~~~\bar M_{12} =
M_{40} \ , \nn \\
&&
\bar M_{30} = -i M_{32} \ , ~~~ \bar M_{31} = -i M_{42} \ , ~~~ \bar
M_{32} = i M_{02}  \ , \nn \\
&&
\bar M_{40} = M_{31} \ , ~~~\bar M_{41} = M_{41} \ , ~~~\bar M_{42} =
- M_{01} \ , ~~~ \bar M_{43} = i M_{12}   \ .
\eea
which in turn results into the following map 
\bea
&&
\bar D = i J_{12} \ , ~~~ \bar J_{01} = D \ , ~~~ \bar J_{02} =
\f{1}{2} \l( P_0-K_0 \r ) \ , ~~~ \bar J_{12}= \f{1}{2} \l( P_0+K_0 \r
) \ ,  \nn \\
&&
\bar P_0 = \f{1}{2} \l [ P_1-K_1 - i \l ( P_2-K_2 \r )\r ] \ , ~~~
\bar K_0 = \f{1}{2} \l [ P_1-K_1 + i \l ( P_2-K_2 \r )\r ] \ ,  \nn \\
&&
\bar P_1 = \f{1}{2} \l [ P_1+K_1 - i \l ( P_2+K_2 \r )\r ] \ , ~~~
\bar K_1 = \f{1}{2} \l [ P_1+K_1 + i \l ( P_2+K_2 \r )\r ] \ , \nn \\
&&
\bar P_2 = - \l ( J_{01}-i J_{02} \r ) \ , ~~~ \bar K_2 = - \l (
J_{01}+i J_{02} \r ) \ . 
\eea
The first thing to note here is that the generators $\bar J_{\m \n}$
are Hermitian. Therefore, if we want to realize the Hilbert space on
the space of fields, or 
in other words
\be
| \Phi \ra  = \Phi(0) | 0 \ra \ ,
\ee
the equation
\be
\l [ \bar J_ {\m \n}, \Phi (0) \r ] =  i \Sigma _ {\m \n} \Phi(0),
\ee
necessitates that $\Phi(0)$ be an infinite-component
field~\cite{Barut:1979we,Carmeli:1970ia,Stoyanov:1968tn,Moffat:1987yj}.
From Eqs.~\REF{eq:CFTRepresentationCartan} we find that its spin is
$h$, therefore  not bound to be half-integer, and at the
same time
\be
\l [ \bar D,  \Phi(0) \r ] = i m \Phi(0) \ .
\ee
The corresponding expressions for the lowering generators \REF{eq:RaisingLoweringGenerators} are given by
\bea
&&
M_1^- =\f{1}{2} \l [ \bar P_1-i\bar P_2 +\bar K_1 - i \bar K_2 \r ]  \
, \nn\\
&&
M_2^- = \f{i}{2} \l [ \bar P_1-i\bar P_2 - \l( \bar K_1 - i \bar K_2
\r ) \r ] \ , \nn \\
&&
M_3^- = \bar J_{01} - i \bar J_{02} \ . 
\eea
Clearly, in this frame the states in the Hilbert space are not given by
primary fields at zero. Instead, the fields should satisfy the
following constraints 
\be
\l( \p_1 - i
\p_2 \r ) \Phi (0) = 0 \ , ~~~ \l [ \bar K_{1} - i \bar K_{2}, \Phi (0) \r ] = 0 \ , ~~~ (\Sigma_{01} - i \Sigma_{02} ) \Phi (0)
= 0 \ . 
\ee

\section{Conformal OPE}
\label{app:OPE}

Here we discuss the constraints conformal invariance imposes on the OPE. We start from the general expression 
\be
\label{eq:OPE_general}
\bar \phi_2 (z_2) \bar \phi_1 (z_1) = \sum_\mc O c_{12\mc O} (z_{2},z_{1}) \bar {\mc O}(z_1)\ ,
\ee
without assuming that the sum runs only over primary fields. 

Let us start with translations. Acting with $\bar P_a$ on both sides of the above, we immediately find
\be
\p_a ^{z_1} c_{12\mc O}(z_2,z_1) + \p_a ^{z_2} c_{12\mc O}(z_2,z_1)=0 \ ,
\ee
meaning that 
\be
c_{12\mc O}(z_2,z_1)=c_{12\mc O}(z_2-z_1) \  .
\ee

For the Lorentz transformations---generated by $\bar J_{ab}$---the expansion~(\ref{eq:OPE_general}) gives\,\footnote{To keep the discussion maximally clear and without loss of generality, in what follows we take $z_2=z$, $z_1=0$.}
\bea
&&\Sigma^1_{ab,\{c\}\{p\}} c_{12\mc O}^{\{p\}\{d\}\{f\}}(z)+\Sigma^2_{ab,\{d\}\{p\}} c_{12\mc O}^{\{c\}\{p\}\{f\}}(z) \nn \\
&&\hspace{2.5cm}+\Sigma^{\mc O}_{ab,\{f\}\{p\}} c_{12\mc O}^{\{c\}\{d\}\{p\}}(z)
+(z_a\p_b-z_b\p_a) c_{12\mc O}^{\{c\}\{d\}\{f\}}(z)=0 \ ,
\eea
where $\{\cdot\}$ stands for (possibly multiple) indices corresponding to the $SO(d)$ representation of the operators. This relation implies that if $\bar\phi_1$, $\bar\phi_2$ and 
$\bar{\mc O}$ are traceless symmetric tensors of ranks $l$, $m$ and $n$, respectively, we get for the function $c_{12\mc O}$
\bea
c_{12\mc O}^{\{c\}\{d\}\{f\}}(z) & = & 
z_{c_1}\dots z_{c_{l}} z_{d_1}\dots z_{d_{m}}  z_{f_1}\dots z_{f_{n}}
A_{l+m+n}(z^2) \nn \\
&&
+\d_{c_1d_1}z_{c_2}\dots z_{c_{l}} z_{d_2}\dots z_{d_{m}}
z_{f_1}\dots z_{f_{n}}  A^{12}_{l+m+n-2}(z^2) \nn \\
&&
+\d_{c_1f_1}z_{c_2}\dots z_{c_{l}} z_{d_1}\dots z_{d_{m}}
z_{f_2}\dots z_{f_{n}}  A^{13}_{l+m+n-2}(z^2)  \nn \\
&&
+\d_{d_1f_1}z_{c_1}\dots z_{c_{l}} z_{d_2}\dots z_{d_{m}}
z_{f_2}\dots z_{f_{n}}  A^{23}_{l+m+n-2}(z^2) + \dots \ ,
\eea
where $\dots$ stand for all other terms that can be obtained from contracting indices belonging to the different sets $\{c\}$, $\{d\}$ and $\{f\}$. 

Similarly, acting with $\bar D$ on the OPE, we obtain
\be
\l (\D_1+\D_2-\D \r ) c_{12\mc O}(z) +z^a\p_a c_{12\mc O}(z) = 0 \ .
\ee
In other words, dilatations fix the coefficient functions to be of the following form 
\be
c_{12\mc O}(z) = |z|^{\D-\D_1-\D_2} F\l (n_a \r ),~~~n_a = \f{z_a}{|z|} \ ,~~~|z| = \sqrt{z_a z^a}\ .
\ee

Using all the constraints we got so far, we can write down the OPE of two primary fields with spins $l_1$ and $l_2$ (i.e. two traceless symmetric tensors of ranks $l_1$ and $l_2$, respectively); this reads
\bea
\bar \phi^{\{a\}_{l_2}}_2(z) \bar \phi^{\{b\}_{l_1}}_1(0)
&=&
\sum_{\mc O} |z|^{\D_{\mc O}-\D_1-\D_2} 
\l[ \lambda^{(l_1+l_2+l_{\mc O})}_{\mc O} n^{a_1}\dots n^{a_{l_2}} n^{b_1}\dots n^{b_{l_1}} n^{c_1}\dots n^{c_{l_{\mc O}}} \r. \nn \\
&+&\l. \lambda^{(l_1+l_2+l_{\mc O}-2)}_{\mc O} \d^{a_1 b_1} n^{a_2}\dots n^{a_{l_1}} n^{b_2}\dots n^{b_{l_2}} n^{c_1}\dots n^{c_{l_{\mc O}}} +\dots\r] \mc O^{\{c\}_{l_{\mc O}}}(0) \ , \nn
\eea
where the sum still runs over all possible operators. 

What remains to be understood is what new information on the OPE we extract once we require that it be consistent with special conformal transformations. It turns out that the contributions of descendants are intrinsically linked to those of the  corresponding primaries. Schematically, for every $\lambda$ (their number can be found in~\cite{Kravchuk:2016qvl}), we get
\bea
&&\bar \phi^{\{a\}_{l_2}}_2(z) \bar \phi^{\{b\}_{l_1}}_1(0)  \nn
=
\sum_{\bar \phi} |z|^{\D_{\mc O}-\D_1-\D_2} \times\\
&&\hspace{.5cm}
\times \l \{
\lambda^{(l_1+l_2+l)}_{12\bar \phi} n^{a_1}\dots n^{b_1}\dots n^{c_1}\dots  
\l [ \phi^{c_1\dots c_l}(0)+ a^{(l_1+l_2+l)} z_c \p_c \phi^{c_1\dots c_l}(0)+\dots \r ]
+\dots \r \}\ .\nn
\eea
Note that owing to conformal symmetry, all the coefficients appearing in front of the descendants are fixed. For the OPE of scalar operators those can be found in~\cite{Ferrara:1971vh}. On the other hand, for operators with non-zero spin we get
\be
a^{(l_1+l_2+l)} = \f{\D_{21}+\D+l_{21}+l}{2(\D+l)}\ ,~~~ \D_{21}=\D_2-\D_1\ ,~~~l_{21}=l_2-l_1 \ .
\ee

\section{Hermitian conjugation \& compatibility  with the conformal algebra}
\label{app:herm_conj}

It is instructive to explicitly show that the way we defined Hermitian conjugation, see Eqs.~(\ref{eq:EScalarConjugation})-(\ref{eq:eq:ETensorConjugation}), does not spoil the action of the conformal algebra on fields. In what follows we work with vectors, for which
\be
\label{eq:app_field}
\bar \phi^\dagger_a(z) = z^{-2\D} I_a^b(z) \bar \phi_b (I z) \ .
\ee
Before moving on, let us present some useful formulas. We note that 
\be
\p_a^{Iz} = z^2I_a^b(z)\p_b^z \ ,
\label{eq:derivativezIz}
\ee
and
\be
\p_c I_{ab} (z) = \f{2}{z^2} \l ( \f{2z_az_bz_c}{z^2} - \d_{ac} z_b-\d_{bc}z_a \r ) \ , 
\ee
which imply
\be
I_{a}^d(z)\p_c I_{bd}(z) = \f{2}{z^2} \l ( z_a \d_{bc} - z_b \d_{ac} \r ) \ ,
\ee
and
\be
z^{-2\D}I_{a}^b(z)\p_c^z\bar \phi_b (I z) = \p_c \bar \phi_a^\dagger(z)+2\D\f{z_c}{z^2}\bar \phi^\dagger _a(z)+\f{2}{z^2}\l ( z_a\d_{bc} - z_b \d_{ac} \r ) \bar \phi^\dagger _b(z) \ .
\label{eq:derivativephidagger}
\ee
Using the above, we now turn to the action of the generators on the vector field; the computations are straightforward but a bit long. 

We start from translations, for which 
\bea
\label{eq:app_transl}
\l [\bar P_a, \bar \phi_b (z)\r ]^\dagger
&&\hspace{-.5cm}\overset{\REF{eq:HermCFT}}{=}
-\l [\bar K_a, \bar \phi^\dagger_b (z)\r ] \nn \\
&&\hspace{-.5cm}\overset{(\ref{eq:GeneratorsFieldsEuclideanAction})}{\underset{(\ref{eq:app_field})}{=}} -i\l[\l( 2\f{z_a z^c}{z^2} -\d^c_a\r)\f{z^{-2\D}}{z^2}I_{b}^d(z)\p_c^{Iz}\bar\phi_d(Iz) +\f{2}{z^2}\l(z_a\D_\phi\d^d_b+z^c \Sigma_{ab,c}^{~~~d}\r)\bar\phi^\dagger_d(z)\r]   \nn \\
&&\hspace{-.5cm}\overset{\REF{eq:derivativezIz}}{=} -i\l[\l( 2\f{z_a z^c}{z^2} -\d^c_a\r)I_c^e(z)z^{-2\D}I_{b}^d(z)\p_e^{z}\bar\phi_d(Iz) +\f{2}{z^2}\l(z_a\D_\phi\d^d_b+z^c \Sigma_{ab,c}^{~~~d}\r)\bar\phi^\dagger_d(z)\r] \nn \\
&&\hspace{-.5cm}\overset{\REF{eq:derivativephidagger}}{=} i\p_a\bar\phi^\dagger_b(z) \ .  
\eea

Moving to Lorentz transformations, it is easy to see that 
\bea
\l [\bar J_{ab}, \bar \phi_c (z)\r ]^\dagger 
&&\hspace{-.5cm}\overset{\REF{eq:HermCFT}}{=}
-\l [ \bar J_{ab}, \bar \phi^\dagger_c (z)\r ]\nn \\
&&\hspace{-.5cm}\overset{(\ref{eq:GeneratorsFieldsEuclideanAction})}{\underset{(\ref{eq:app_field})}{=}} -i\Sigma_{ab,c}^{~~~d}\bar\phi^\dagger_d(z) -i\f{z^{-2\D}}{z^2}\l(z_aI_b^d(z)-z_bI_a^d(z)\r)\p_d^{Iz}\bar\phi_c(Iz) \nn \\
&&\hspace{-.5cm}\overset{\REF{eq:derivativezIz}}{=}-i\Sigma_{ab,c}^{~~~d}\bar\phi^\dagger_d(z)-i z^{-2\D}\l(z_aI_b^d(z)-z_bI_a^d(z)\r)I_c^e(z)\p_e^z\bar\phi_c(Iz)\nn \\
&&\hspace{-.5cm}\overset{\REF{eq:derivativephidagger}}{=}-i\l(\Sigma_{ab,c}^{~~~d} +\l(z_a\p_b-z_b\p_a\r)\d^d_c\r)\bar\phi^\dagger_d(z) \ ,
\eea
where the spin matrices $\Sigma_{ab,cd}$ for vectors are defined in~(\ref{eq:vec_rep_spin}). 

For dilatations, a completely analogous computation shows that 
\bea
\l [\bar D, \bar \phi_b (z)\r ]^\dagger 
&&\hspace{-.5cm}\overset{\REF{eq:HermCFT}}{=}
\l [\bar D, \bar \phi^\dagger_b (z)\r ]\nn \\ 
&&\hspace{-.5cm}\overset{(\ref{eq:GeneratorsFieldsEuclideanAction})}{\underset{(\ref{eq:app_field})}{=}}-i\D\bar\phi^\dagger_b(z)-i \f {z^a}{z^2}z^{-2\D}I^c_b(z)\p_a^{Iz}\bar\phi_c(Iz) 
\nn \\
&&\hspace{-.5cm}\overset{\REF{eq:derivativezIz}}{=} -i\D\bar\phi^\dagger_b(z) - i z^a z^{-2\D}I^c_b(z)I^d_a(z)\p_d^z\bar\phi^\dagger_c(z)\nn\\
&&\hspace{-.5cm}\overset{\REF{eq:derivativephidagger}}{=}
i \l(\Delta + z^a\p_a\r)\bar\phi^\dagger_b(z) \ .  
\eea

Finally, the action of special conformal transformations reads
\bea
\label{eq:app_SCTs}
\l [ K_a, \bar \phi_b (z)\r ]^\dagger 
&&\hspace{-.5cm}\overset{\REF{eq:HermCFT}}{=}
-\l [ P_a, \bar \phi^\dagger_b (z)\r ]\nn\\
&&\hspace{-.5cm}
\overset{(\ref{eq:GeneratorsFieldsEuclideanAction})}{\underset{(\ref{eq:app_field})}{=}}
iz^{-2\D}I_{bc}(z)\p_a^{Iz}\bar\phi_c(Iz) \nn\\
&&\hspace{-.5cm}
\overset{\REF{eq:derivativezIz}}{=}
iz^2I_{ad}(z)z^{-2\D}I_{bc}(z)\p_d^z\bar\phi_c(Iz) \nn \\
&&\hspace{-.5cm}\overset{\REF{eq:derivativephidagger}}{=}
-i  \Big [ \l ( 2z_a z^c-z^2\d_{a}^c\r )\d^d_b\p_c  +2\l ( z_a \D \d_{b}^d +z^c \Sigma_{ac,b}^{~~~d} \r )\Big ] \bar \phi^\dagger_d (z)  \ . 
\eea

Inspection of the commutators~(\ref{eq:app_transl})-(\ref{eq:app_SCTs})  reveals that they are indeed  consistent with~\REF{eq:GeneratorsFieldsEuclideanAction}.

\section{Schr\"odinger algebra}
\label{app:schro_al}

In a $d$ dimensional spacetime the Schr\"odinger algebra satisfies the following commutation relations\,\footnote{ The commutator of $P$ and $K$ is obtained by central extension.}
\bea
\l [ J _ {i j}, J _ {k l} \r ] & = & i \l ( J _ {j l} \d _ {i k} + J _
{i k} \d _ {j l} - J _ {i l} \d _ {j k} - J _ {j k} \d _ {i l}  \r ) \
, \nn \\
\l [ J _ {i j}, P _ {k} \r ] & = & i \l ( \d _ {i k} P _ j - \d _ {j
  k} P _ i  \r ) \ , \nn \\
\l [ J _ {i j}, K _ {k} \r ] & = & i \l ( \d _ {i k} K _ j - \d _ {j
  k} K _ i  \r ) \ , \nn \\
\l [ K _ {i}, P _ {j} \r ] & = & - i \d _ {i j} Q \ , \nn \\
\l [ K _ {i}, H \r ] & = & - i P _ i \ , \nn \\
\l [ D, H \r ] & = & - 2 i H\ , \nn \\
\l [ D, P _ i \r ] & = & - i P _ i\ , \nn \\
\l [ D, K _ i \r ] & = & i K _ i\ , \nn \\
\l [ D, S \r ] & = & 2 i S\ , \nn \\
\l [ S, H \r ] & = & i D\ , \nn \\
\l [ S, P _ i \r ] & = & i K _ i\ . 
\label{nrcft}
\eea

In what follows we present the expressions for the  generators of the Schr\"odinger group away from the origin. These are useful for obtaining the action of the algebra on fields, see~(\ref{eq:FieldsNRCFT}).
In deriving them, we used the commutation relations presented above and Baker-Campbell-Hausdorff formula for two operators $A$ and $B$
\be
\label{eq:BCH_form}
e^{-A}B e^A = B +[B,A]+\f{1}{2!} [[B,A],A] +\ldots \ . 
\ee
\begin{itemize}
    \item Translations (we denote $P y = H y_0 - y_i P_i$)
\bea
\label{eq:app_transl_action}
e ^ {- i P y} J _ {i j} e ^ {i P y} & = & J _ {i j} + y_i P _ j - y _
j P _ i\ , \nn \\
e ^ {- i P y} K _ i e ^ {i P y} & = & K _ i + y _ 0 P _ i - y _ i Q \ , \nn \\
e ^ {- i P y} D e ^ {i P y} & = & D + 2 y _ 0 H - y _ i P _ i \ , \nn \\
e ^ {- i P y} S e ^ {i P y} & = & S - y _ 0 D + y _ i K _ i - y _ 0 ^
2 H + y _ 0 y _ i P _ i - \f {1} {2} y _ i^2 Q\ .
\eea
\item Angular momentum\,\footnote{We define the rotation matrix $R_{ij}$ as the vector representation of the rotation group
\be
R_{ij} = \rho_{vec}(e^{-iJ_{ij}\a_{ij}/2}) \ .
\ee}
\bea
 e ^ {-i J_{kl} \a_{kl} / 2} P _ i e ^ {i J_{kl} \a_{kl} / 2} & = & P _ j R _ {j i}\ , \nn \\
 e ^ {-i J_{kl} \a_{kl} / 2}J _ {i j}  e ^ {i J_{kl} \a_{kl} / 2}& = & J _ {k l} R _ {k
  i} R _ {l j}\ , \nn \\
 e ^ {-i J_{kl} \a_{kl} / 2} K _ i  e ^ {i J_{kl} \a_{kl} / 2}& = & K _ j R _ {j i} \ .
\eea
\item Boosts ($\vec K \vec a = K_i a_i$)
\bea
e ^ {- i \vec K \vec a}  H e ^ {i \vec K \vec a} & = & H - a _ i P _ i
+ \f {1} {2} a _i^2 Q\ , \nn \\
e ^ {- i \vec K \vec a}  P _ i e ^ {i \vec K \vec a} & = & P _ i - a _
i Q \ , \nn \\
e ^ {- i \vec K \vec a} J _ {i j} e ^ {i \vec K \vec a} & = & J _ {i
  j} + K _ i a _ j - K _ j a _ i\ , \nn \\
e ^ {- i \vec K \vec a} D e ^ {i \vec K \vec a} & = & D - a _ i K _ i
\ .
\eea
\item Dilatations
\bea
e ^ {- i D \a} H e ^ {i D \a} & = & H e ^ {- 2 \a} \ , \nn \\
e ^ {- i D \a} P _ i e ^ {i D \a} & = & P _ i e ^ {-\a}\ , \nn \\
e ^ {- i D \a} K _ i e ^ {i D \a} & = & K _ i e ^ {\a}\ , \nn \\
e ^ {- i D \a} S e ^ {i D \a} & = & S e ^ {2 \a}\ ,
\eea
\item Special conformal transformations
\bea
e ^ {- i S \a} H e ^ {i S \a} & = & H + \a D - \a ^ 2 S \ , \nn \\
e ^ {- i S \a} P _ i e ^ {i S \a} & = & P _ i + \a K _ i \ , \nn \\
e ^ {- i S \a} D e ^ {i S \a} & = & D - 2 \a S \ .
\eea
\end{itemize}

\section{Coordinate transformation \label{app:NRbarTransform}}

To derive the automorphism relating the two frames, we used 
\bea
e^{-i (H-S)\a} P_i e^{i (H- S)\a} & = & P_i \cos \a - K_i \sin \a \ , \nn \\
e^{-i (H-S)\a} K_i e^{i (H- S)\a} & = & P_i \sin \a + K_i \cos \a\ , \nn \\
e^{-i (H-S)\a} S e^{i (H- S)\a} & = & S \cos^2 \a -H \sin^2 \a -\f{1}
{2}\sin 2 \a D \ , \nn \\
e^{-i (H-S)\a} H e^{i (H- S)\a}& = & H \cos^2 \a -S \sin^2 \a -\f{1}
{2} \sin 2 \a D \ , \nn \\
e^{-i (H-S)\a} D e^{i (H- S)\a} & = & D \cos 2 \a + (H+S) \sin 2 \a \ . 
\eea

\section{NRCFT field transformations \label{app:NRCFTGeneralTransformation}}

Here we discuss how a primary field behaves under an arbitrary Schr\"odinger transformation 
\be
\phi(t, \vec x) \equiv \bar g \phi'(t, \vec x) \bar g^{-1} \ ,
\label{eq:TransformationGeneric}
\ee
with $\bar g$ belonging to the Schr\"odinger group. We note that we can always write
\be
\bar g = g e^{i \omega J/2} \ .
\ee
Since the action of rotations is obvious, we can focus on transformations that do not involve~$J$. 

As usual, in order to find an explicit expression for~(\ref{eq:TransformationGeneric}),  we first consider the action of the element $g$ on the  coordinates, namely
\be
\Omega\equiv g e^{iHt} e^{-i\vec P \vec x} \ ,
\label{eq:groupLeftAction}
\ee
and rewrite it as the action on the coset space
\be
\Omega = e^{iHt'} e^{-i\vec P \vec x' } e^{-i \s D} e^{-i \vec \b \vec
  K} e^{i \rho S} e^{i \psi Q} \ ,
\label{eq:CosetAction}
\ee
with new coordinates $\l(t'=t' (t,\vec x),~\vec x'=\vec x'(t,\vec
x)\r)$ and the yet to be derived 
parameters $\s=\s (t,\vec x)$, $\vec \b=\vec \b (t,\vec x)$, $\rho=\rho (t,\vec x)$ and $\psi=\psi (t,\vec x)$. Equating the Maurer-Cartan forms $\Omega^{-1}\p_\m \Omega$ for both~(\ref{eq:groupLeftAction}) and~(\ref{eq:CosetAction}), we get 
\bea
H \d_{\m t} - P_i \d_{\m i} & = & H e^{2\s}\p_\m t' - P_i \l ( e^\s \p_\m x_i' -e^{2\s} \p_\m t' \b_i \r ) -D \l ( \p_\m \s - e^{2\s}\p_\m t' \rho \r ) \nn \\
&&
-K_i \l [ \p_\m \b_i +\b_i \p_\m \s +\rho \l ( e^\s \p_\m x_i' -e^{2\s} \p_\m t' \b_i \r ) \r ] \\
&&
+ S \l ( \p_\m \rho - \rho^2 e^{2\s}\p_\m t'  +2 \rho \p_\m \s \r )+Q \l ( \p _\m \psi +\f{1}{2} \vec \b^2 e^{2\s}\p_\m t'  - e^\s \p_\m x_i' \b_i \r ) \ . \nn
\eea
Comparing the coefficients of the various generators in the above allows to express the parameters $\s$, $\vec \b$, $\rho$ and $\psi$ in terms of transformed coordinates. 

We start from $H$, for which we first observe that  
$\p_i t' =0$, and also
\be
\s = -\f 1 2 \log \p_t t' \ .
\ee
Moving to $P_i$, for the spatial derivative we find
\be
e^\s \p_j x_i' = \d_{ij} \ ,
\ee
which implies that
\be
 x_i' = x_i \sqrt{\p_t t'}+g_i(t) \ ,
\ee
with $g_i(t)$ being an arbitrary function of $t$. For the time derivative we obtain
\be
\b_i = \f{ x_i}{2}\f{\p_t^2 t'}{\p_t t'}+\f{\p_t g_i}{\sqrt{\p_t t'}}
\ .
\ee
Similarly, from the coefficients of  $D$ and $Q$ we respectively get
\be
\rho = -\f{1}{2}\f{\p_t^2 t'}{\p_t t'} \ ,
\ee
and
\be
\p_i \psi = \b_i= \f{x_i}{2}\f{\p_t^2 t'}{\p_t t'}+\f{\p_t
  g_i}{\sqrt{\p_t t'}}\ ,~~~ \p_t \psi = \f{\b_i^2}{2} =\f 1 2\l (  \f{
  x_i}{2}\f{\p_t^2 t'}{\p_t t'}+\f{\p_t g_i}{\sqrt{\p_t t'}} \r )^2 \ .
\label{eq:Psi Constraint}
\ee
The latter two relations are consistent provided that
\be
\p_t\b_i = \b_j \p_i \b_ j \ ,
\ee
which translates into the  Schwarzian derivative of $t'$ vanishing
\be
(S t') (t) \equiv \f{\p^3_t t'}{\p_t t'} - \f{3}{2}\l ( \f{\p^2_t
  t'}{\p_t t'} \r )^2 = 0 \ ,
\label{eq:SchwarziaT}
\ee
and at the same time $g_i(t)$ being subject to
\be
\label{eq:constr_gi}
\p_t^2 g_i = \p_t g_i \f{\p_t^2 t'}{\p_t t'} \ .
\ee
The solution to \REF{eq:SchwarziaT} is  an arbitrary M\"obius transformation
\be
t' = \f{at+b}{ct+d} \ ,
\ee
while \REF{eq:constr_gi} fixes $g_i$ to be a linear function of $t'$
\be
g_i(t) = v_i t'(t)+a_i \ ,
\ee
with $v_i$ and $a_i$ arbitrary constants.

Knowing that, Eq.~(\ref{eq:Psi Constraint}) can be integrated to produce
\be
\psi = \f{x_i^2}{4}\f{\p_t^2 t'}{\p_t t'}+ v_i  x_i \sqrt{\p_t t'}
+\f{ v_i^2 t'}{2}+\a \ , ~~ \a=\const \ .
\ee

Collecting everything together, we conclude that
\bea
\phi(t, \vec x) & = & \Omega \phi'(0) \Omega^{-1} = 
e^{iHt'} e^{-i\vec P \vec x'} e^{-i \s D} e^{i \psi Q} \phi' (0) e^{-i
  \psi Q} e^{i \s D}  e^{i\vec P \vec x'} e^{-iHt'} \nn \\
& = &
e^{-i m \psi} e^{-\D \s}e^{iHt'} e^{-i\vec P \vec x'} \phi' (0) e^{i\vec P \vec x'} e^{-iHt'} =
e^{-i m \psi} e^{-\D \s} \phi' (t',\vec x') \nn \\
& = &
\l ( \p_t t' \r )^{\D/2} \exp\l[-im\l(\f{x_i^2}{4}\f{\p_t^2 t'}{\p_t
  t'}+ v_i  x_i \sqrt{\p_t t'} +\f{ v_i^2 t'}{2}+\a\r)\r] \phi'
(t',\vec x')  \ .
\label{eq:NRCFTGeneralTransformation}
\eea

\section{Non-trivial geometry \label{app:NewtonCartan}}

In this section we quickly recap how non-relativistic systems can be
coupled to a non-trivial background  geometry. This can be done using
the so-called coset
construction~\cite{Delacretaz:2014oxa,Karananas:2016hrm}. Considering
the Galilei group $\mathrm{Gal}$, we introduce the following coset representative
\be
\Omega = e^{i H w_0(x)}e^{-i P_i w^i (x)} \ .
\ee 
The Maurer-Cartan form is given by
\be
\Theta _\m = -i \Omega^{-1} \tilde D_{\m} \Omega \equiv \Omega^{-1} \l( \p_\m + i \tilde n_\m H - i \tilde e^i _\m P_i + i \omega_\m^i K_i 
+\f{i}{2} \t^{ij}_\m J_{ij} + i \tilde A_\m Q \r ) \Omega \ ,
\label{eq:MCformGravity}
\ee
where $\tilde n_\m$, $\tilde e^i_\m$, $\omega^i_\m$, $\t^{ij}_\m$ and
$\tilde A_\m$ are gauge fields corresponding to time and space
translations, boosts, spatial and $U(1)$ phase  rotations,
respectively. Their transformation properties are meant precisely for
canceling the left action of the group, i.e. for $g \in \mathrm {Gal}$
we demand that\,\footnote{The gauge fields are collectively denoted by
  $X_\m$. }
\be
\Omega^{-1} g^{-1} \l ( \p_\m + i X'_\m \r ) g \Omega = \Omega^{-1} \l
( \p_\m + i X_\m \r ) \Omega \ ,
\ee
leading to
\be
X_\m' = g X_\m g^{-1}+ i \p_\m g g^{-1} \ .
\ee
Note that neither $\tilde n_\m$ or $\tilde e^i _\m$ transform as
vielbeins. In order to get the latter, the auxiliary fields $w_0(x)$
and $\vec w(x)$ should be absorbed into the new definitions of $n_\m$, $e^i _\m$ and $A_\m$. Simplifying \REF{eq:MCformGravity} we get
\be
\Theta _\m = n_\m H - e^i _\m P_i + Z_\m \ ,
\label{eq:MCformGravitySimplified}
\ee
with
\be
Z_\m= \omega_\m^i K_i +\f{1}{2} \t^{ij}_\m J_{ij} +A_\m Q \ .
\ee
The fields $n_\m$ and $e_\m^i$ transform as temporal and spatial
vielbeins, while the transformation of $Z_\m$ is that of a gauge
field. Explicitly, we can deduce the  transformations of all fields from the standard transformations of the coset. Namely, for
\be
g \Omega = \Omega' h \ ,
\ee
with $h$ being an element of a subgroup of $\mathrm{Gal}$ generated by rotations, boosts and $U(1)$ transformations, which we denote by 
$\mathrm{Gal} \backslash \{ H, \vec P\}$, we get
\be
\Theta_\m ' \equiv -i (\Omega')^{-1} D'_\m \Omega' = h \Theta h^{-1}+i
\p_\m h h^{-1} \ ,
\ee
while for matter fields $\phi(t,\vec x)$ belonging to a representation $\rho$ of $\mathrm{Gal} \backslash \{ H, \vec P\}$ (which can be read off the commutation relations $\rho (X) \phi(0) = - \l [ X, \phi (0) \r ]$), we obtain 
\be
\phi'(x) = \rho (h) \phi(x) \ .
\ee
The covariant derivatives of matter fields are given by
\be
D_\m \phi = \p_\m \phi + i \rho (Z_\m) \phi \ .
\ee

In Table~\ref{tab:TransformationsNewtonCartan} we present the
transformations of the geometric data and matter fields under
rotations, boosts and $U(1)$ phase rotations with
parameters $\a_{ij}$, $v_i$ and $\a$, respectively: 
\be
\label{tab:TransformationsNewtonCartan}
\bt{c | c c c}
 & $e^{-iJ_{ij}\a_{ij}/2}$ & $e^{-i \vec K \vec v} $& $e^{-i Q\a}$\\
\hline
$n ' _ {\m}$ & $n _ \m$ & $n _ \m$ & $n _ \m$ \\
$e ^ {'i} _ \m$ &  $ R_{j}^i e^j _\m$ & $e _ \m ^ i + v ^ i n _ \m$ &  $e^i _  \m$ \\
$A _ \m '$ & $A_\m$ & $A _ \m + v_i e^i_\m+\f 1 2 \vec v^2 n_\m $ &  $A _ \m + \p _ \m \a$ \\
$\phi '$ & $\rho(R) \phi$ & $\phi $ &  $\phi e^{- i m \a}$
\et
\ee
It should be noted~\cite{Karananas:2016hrm} that  under boosts  the actual 
transformation of the matter and gauge fields $\phi$ and $A_\m$  contains an additional 
$U(1)$ rotation, which is a pure gauge transformation, therefore, it was dropped. 

\bibliographystyle{utphys}
\bibliography{NRCFT}{}

\end{document}